\documentclass[aps,twocolumn,amsmath,amssymb,floatfix]{revtex4}

\usepackage{amsmath,amsthm}
\usepackage{graphicx}% Include figure files
\usepackage{dcolumn}% Align table columns on decimal point
\usepackage{bm}% bold math
\usepackage{color}
\usepackage{epsfig}
\usepackage{multirow}
\usepackage{mathrsfs}
\usepackage{hyperref}
\usepackage{cleveref}
\usepackage{epstopdf}

\begin{document}

\preprint{Preprint}

\title{Self-consistent field model for strong electrostatic correlations and \\ inhomogeneous dielectric media}

\author{Manman Ma}\email{mmm@sjtu.edu.cn}
\author{Zhenli Xu}\email{xuzl@sjtu.edu.cn}
\affiliation{Department of Mathematics, Institute of Natural Sciences, and MoE Key Lab of Scientific and Engineering Computing, Shanghai Jiao Tong University, Shanghai 200240, China}


\date{\today}

%%%%% Begin Abstract %%%%%%%%%%%
\begin{abstract}
Electrostatic correlations and variable permittivity of electrolytes are essential for exploring many chemical and physical properties of interfaces in aqueous solutions. We propose a continuum electrostatic model for the treatment of these effects in the framework of the self-consistent field theory. The model incorporates a space-
or field-dependent dielectric permittivity and an excluded ion-size effect for the correlation energy. This results in a self-energy modified Poisson-Nernst-Planck or Poisson-Boltzmann equation together with state equations for the self energy and the dielectric function. We show that the ionic size is of significant importance in predicting a finite self energy for an ion  in an inhomogeneous medium. Asymptotic approximation is proposed for the solution of a generalized Debye-H\"uckel equation, which has been shown to capture the ionic correlation and dielectric self energy. Through simulating ionic distribution surrounding a macroion, the modified self-consistent field model is shown to agree with particle-based Monte Carlo simulations. Numerical results for symmetric and asymmetric electrolytes demonstrate that the model is able to predict the charge inversion at high correlation regime in the presence of multivalent interfacial ions which is beyond the mean-field theory, and also show strong effect to double layer structure due to the space- or field-dependent dielectric permittivity.
	
\end{abstract}
%%%%% end %%%%%%%%%%%

%%%%% AMS/PACs/Keywords %%%%%%%%%%%

%\pacs{02.70.Uu, 83.10.Rs, 82.70.Dd, 61.20.Ja}
%\ams{}
%\keywords{}

%%%% maketitle %%%%%
\maketitle

 %%%% Start %%%%%%
\section{Introduction}

Electrostatic interaction in aqueous solutions
including charged interfaces is of interest in a wide range of areas such as biological macromolecules, colloidal suspensions and nanoparticle assembly. The structure of screened ions near charged surfaces, so-called the electric double layer, controls many macroscopic properties of the system such as the zeta potential and colloidal renormalized charges \cite{Levin:RPP:2002,Hunter::1981,JYTB:PRL:04,FPP+:RMP:2010,WK+:N:2011}, and is affected by the interfacial chemistry, the surface charge distribution, the ionic specificity, and so on. In the presence of divalent ions, electrostatic correlation plays a very important role, which could lead to the overscreening of counterions near highly charged surfaces, i.e., the charge inversion phenomenon \cite{BZH+:PRL:2004,PBV+:PRL:2006,LSH:JCP:82,DJH+:JPCB:2001,GNS:RMP:2002}. The electrostatic correlation is probably responsible for a lot of many-body phenomena such as like-charge attraction \cite{LG:N:1997,ALW+:PNASU:2003} and ion crowding in membrane channels \cite{Eisenberg:ACPip:2011a}, and many-body systems such as electrodes in ionic liquids \cite{FK:CR:14}. The correct understanding of correlation effects and their
relation with interfacial properties are then considered to be of significant in many applications.

Electrostatic correlation
should be seriously studied in many conditions, e.g., if a system includes highly charged
surfaces, multivalent counterions, or if the system is at low temperature. The correlation strength is often measured by a coupling parameter \cite{BKN+:PR:2005} defined as $\Xi= z_c^3 \sigma e^4/8\pi(\varepsilon k_B T)^2,$ where $z_c$, $\sigma$, $e$, $\varepsilon$ and $k_BT$ are the counterion valence, the average surface charge density, the electronic charge, the dielectric permittivity, and the thermal energy, respectively. The coupling parameter can be simplified to  $\Xi=z_c^2\ell_B/\ell_{GC}$, the product of $z_c^2$ and the ratio of the Bjerrum length $\ell_B$ to the Gouy-Chapman length $\ell_{GC}$. These two length scales describe the ion-ion and ion-interface interaction strengths, respectively. The classical Poisson-Boltzmann (PB) theory is often used, but it is mean-field and works only when $\Xi\ll 1$ since it ignores the ion correlation, thus fails to capture correlation-induced electrostatic phenomena. Many extended theories beyond the mean field
have been proposed and applied for different problems, e.g., to include the steric effect or dielectric self energy \cite{BAO:PRL:1997,KBA:PRE:2007,CKC:BJ:03,GKCN:JPC:04,EHL:JCP:2010}. A first-principle description of electrostatic correlation should follow the integral-equation theory \cite{HM::2006}, which turns out to be a difficulty in resolving high-dimensional obstacles. Phenomenologically, the electrostatic correlation is taken into account in a Ginzburg-Laudau-type functional which yields a modified PB equation with a biharmonic term to describe the contribution from the correlation energy \cite{BSK:PRL:2011,LE:JPCB:13} and a correlation length as a parameter to be accurately fitted by simulations or experiments.

In most of theoretical and computational studies under the primitive model \cite{Linse:APS:2005,Messina:JPCM:2009}, a homogeneous dielectric profile for electrolytes near surfaces is often assumed, e.g., 80 for the relative dielectric constant of water solvent. This is far from a high-accurate description of the solvent.  Water molecules are ordered near micro- or macro- ions, weakening the shielding for ion-ion interactions. This has led to models for ion-concentration- or field-dependent dielectric permittivities. Historically, the dependence of the dielectric permittivity on the ionic solution has been discussed as early as 1940s \cite{HRC:JCP:48}. Later, Booth considered the dielectric constant of polar electrolytes as a function of the electrostatic field strength \cite{B:JCP:1951, B:JCP:1955}, which was followed by many related studies from molecular dynamics simulations \cite{YB:JCP:99, Fulton:JCP:09} to continuum theory \cite{GRB:JCIS:78, PDKBM:JCIS:96}. Recent work also reported that the dielectric decrement is relevant to ion-specific effects \cite{BAP:JCP:11}.
On the other hand, the Booth's model was utilized to modify the Poisson-Boltzmann equation to predict the capacitance with effects of morphology and dielectric permittivity \cite{WVP:EA:11}. 
Coupled with the Booth's model, the Langevin-Poisson-Boltzmann (LPB) equation was developed to investigate the effects of polarization saturation to the contact potential at the colloid surface and the long-range interaction between colloids \cite{FO:PCCP:11}. More recently, Bonthuis \textit{et al}. \cite{BGN:PRL:2011, BGN:Langmuir:12} and Bonthuis and Netz \cite{BN:JPC:13} through molecular dynamics simulations showed that the interfacial dielectric function is closely related to the molecular solvent structure of the surface. By incorporating the structure properties in the dielectric profile, the Bonthuis-Netz model \cite{BN:JPC:13} has been successful in predicting the ion distribution with good agreement to a bunch of physical experimental results and particle simulations.

When moving an ion from a high-dielectric region to a lower one, it costs the solvation energy and thus favors to stay away. This explains the ion depletion near the water/air interface \cite{OS:JCP:1934}.  So with the decrement of the double-layer dielectric permittivity by the field, ions are repelled. In contrast, we should see that the ion-ion correlation becomes stronger for lower $\varepsilon$ since the coupling parameter $\Xi$ is proportional to $1/\varepsilon^2$, favorable for the ion congregation. These mutual effects to the interfacial ion structure are less understood. This is our purpose to find an appropriate model to account for both effects. We considered the effect of inhomogeneous dielectric function, which is either space-dependent or field-dependent under the continuum theory. We go one step further to couple the dielectric function with the self-consistent field (SCF) theory \cite{podgornik1989jcp, NO:EPJE:2000, NO:EPJE:2003, Wang:PRE:2010,WangRui:JCP:13, LX:PRE:2014,BAA:JCP:2012,LLP+:PRE:2002} to incorporate the correlation energy (the self energy of a test ion). A self-consistent system composed of a self-energy modified PB equation (or Possion-Nernst-Planck equation for charge dynamics) and a generalized Debye-H\"uckel (DH) equation is derived, which can characterize the inhomogeneous dielectric background. This SCF model gives rise to many difficult issues for numerical approximation. On one hand, the inhomogeneity of dielectric permittivity introduces the difference in the solvation energy (Born energy), which is inversely proportional to the product of the dielectric constant and the ionic radius \cite{Born:ZP:1920}. It is necessary to handle the exclude-volume effect of ions, essentially giving rise to a multi-scale problem, which was discussed by Wang \cite{Wang:PRE:2010}.
On the other hand, the solution of the generalized DH equation is always a difficult issue even when the dielectric profile is homogeneous, since it is a Green's function equation. Numerical methods for this high-dimensional problem are computationally expensive. More efficient
numerical methods should be proposed. These technical issues will be resolved by introducing an asymptotic decomposition of the self energy, where the approximate DH equation is represented by an integral over the three-dimensional domain.

The traditional SCF equations for charged systems are based on a point-charge assumption, formulated from the variational field theory. The solution of this set of equations blows up at a high coupling parameter $\Xi$ because the
ionic size effect is ignored and thus the correlation energy is divergent \cite{XuMaggs:JCP:14}. This shortcoming is fixed in this paper, thus allowing us to explore new physics at the higher coupling regime. we simulate
%the
1:1 and 2:1 salts for electrolytes through the proposed SCF model and particle-based Monte Carlo simulations, and find excellent agreement between two approaches. Meanwhile, it is illustrated that many important features are captured by the modified SCF equations, e.g., the depletion due to the low dielectric zone and the charge inversion for strong-coupling systems. Furthermore, we find from the Booth dielectric model that the electric field causes the ordered water alignment only for the first few layers, which however greatly changes the surface potential of the charged interface. These results demonstrate attractive features of the modified SCF model, and then promising for the use in understanding more physical and biological phenomena.

\section{Model}

\subsection{Self-consistent field model with variable permittivity}

Let $\varepsilon$ be the dielectric permittivity of the solvent with a binary salt of valences $z_\pm$. Subject to suitable boundary conditions, the electrostatic potential $\Phi$ in the electrolyte is described by the classical Poisson equation,
\begin{equation}
	-\nabla\cdot\varepsilon\nabla\Phi = \sum_{i=\pm} z_iec_i,
\end{equation}
where $e$ is the elementary charge and $c_i$ is the concentration function of the ions of species $i$.	
By the mass conservation, the modified Nernst-Planck equation is used to describe the dynamics of charged particles in an electrolyte, i.e., the equation for $c_i$ is,
\begin{equation}\left\{\begin{array}{ll}
\displaystyle \frac{\partial c_i}{\partial t} = \nabla\cdot D_i\left(\nabla c_i+\beta c_i\nabla U_i \right), \label{eq:NP}\\
\displaystyle U_i = z_ie\Phi + \frac{1}{2}z_i^2e^2u_i,
\end{array}\right.
\end{equation}
where $D_i$ is the diffusion constant, and $\beta=1/k_BT$ is the inverse thermal energy. The first term in the right side of Eq. \eqref{eq:NP} describes the diffusion of ions, and the second term is the contribution from the ion convection due to the energy gradient and $U_i$ is the electrostatic energy of an ion at the position, called the potential of mean force as approximated by the sum of the mean potential energy and the correlation energy $\frac{1}{2}z_i^2e^2u_i$, following a lot of work \cite{Luo06,NO:EPJE:2003,NO:EPJE:2000,LX:PRE:2014}. The physical meaning of the potential of mean force is the free energy cost of moving a test ion from the bulk solvent region into its current position, where the correlation energy is related to a Green's function to be discussed later on. In equilibrium, the Nernst-Planck equation Eq. \eqref{eq:NP} has an explicit formula for ionic concentration,
$	c_i = c_{i0}e^{-\beta U_i},$
where $c_{i0}$ is the concentration in bulk solution.
This leads to the self-energy modified PB equation,
\begin{equation}
	-\nabla\cdot\varepsilon\nabla\Phi = \sum_i z_iec_{i0}e^{-\beta U_i}. \label{eq:SEMPB1}
\end{equation}
This equation is beyond the classical PB theory in the sense that electrostatic correlations have been included in the self energy to improve the approximation of the mean-force potential.

Equation \eqref{eq:SEMPB1} is available for cases with symmetric or asymmetric ionic sizes and valences.
For physical systems with high surface charge densities or large surface electric potentials, since the ion density near the interface increases exponentially and does not saturate. We introduce a local treatment by using the entropic contribution of water molecules in the free energy and use the lattice gas formalism to derive the ion density distribution with maximal charge density constraints  \cite{BAO:PRL:1997}, though the influence of this treatment is debated in \cite{Gillespie:MN:14}.
In this case, the equilibrium ion concentration with steric entropic effect of each particle in the solvent is given by,
$	c_i = c_{i0}e^{-\beta U_i-S}, $
with
\begin{equation}
S=\log \big[1+\sum_j \nu c_{j0}\left( e^{-\beta U_j}-1 \right)\big],
\end{equation}
where $\nu$ is the lattice volume of an ion. At the dilute limit or at the bulk solvent $S$ decays to zero. We then  modify Eq. \eqref{eq:SEMPB1} into,
\begin{equation}
	-\nabla\cdot\varepsilon\nabla\Phi =  \sum_i z_iec_{i0}e^{-\beta U_i-S}. \label{eq:SEMPB2}
\end{equation}
It should be noted that the use of asymmetric lattice volumes is also possible though the relation between the ion concentration and the electric energy has to be in a form of a transcendental equation \cite{Li:N:2009,LLXS:NonL:2013}, which has
the exact explicit solution only for the same size. We should remark that the ionic steric effect is essentially short-range correlation effect and nonlocal \cite{FL:JCP:12}. This should be modified by, e.g., the modified fundamental measure theory \cite{rosenfeld:PRL:1989}, when this steric effect has to treated more accurately. 

In order to close the above equation system, we define the self energy $u_i$ by following the SCF theory \cite{NO:EPJE:2000,NO:EPJE:2003,Wang:PRE:2010}, where it is defined as the self-Green's function limit,
\begin{equation}
	u_i = \lim_{\mathbf{r}'\rightarrow\mathbf{r}}\big[G_i(\mathbf{r},	\mathbf{r}')-G_{0}(\mathbf{r},\mathbf{r}')\big], \label{ugfs}
\end{equation}
and $G_i$ is the Green's function described by a generalized Debye-H\"uckel (DH) equation, and $G_0$ is the free-space Green's function described by $- \varepsilon_\textup{eff} \nabla^2 G_0(\mathbf{r},\mathbf{r}') =\delta(\mathbf{r},\mathbf{r}')$. In order to take into account the effect of variable dielectric permittivity, the DH equation in the SCF theory is expressed as \cite{XML:PRE:2014},
\begin{equation}
-\nabla \cdot \varepsilon_i(\mathbf{r},\mathbf{r}') \nabla G_i(\mathbf{r},\mathbf{r}') +2I_i(\mathbf{r},\mathbf{r}') G_i(\mathbf{r},\mathbf{r}')=\delta(\mathbf{r},\mathbf{r}'), \label{gifs}
\end{equation}
where the dielectric permittivity $\varepsilon_i(\mathbf{r},\mathbf{r}')$ and ionic strength $I_i(\mathbf{r},\mathbf{r}')$ locally depend on the position of the test ion of species $i$, which characterizes its ionic size effects to the dielectric function and excluded volume,
\begin{eqnarray}
    \varepsilon_i(\mathbf{r},\mathbf{r}') &&= \left\{ \begin{array}{ll}
%                \varepsilon_0,~~~~~~~~~~~~~~~|\mathbf{r}-\mathbf{r}'|<a_i,\\
		\varepsilon_\textup{eff},~~~~~~~~~~~~~~~|\mathbf{r}-\mathbf{r}'|<a_i,\\
                \varepsilon (\mathbf{r}),~~~~~~~~~~~~~\hbox{otherwise},
\end{array}\right. \label{eq:ei}\\
    I_i(\mathbf{r},\mathbf{r}') &&= \left\{ \begin{array}{ll}
                0,~~~~~~~~~~~~~~~~~|\mathbf{r}-\mathbf{r}'|<a_i,\\
                \frac{1}{2}\beta e^2\sum_i z_i^2c_i,~~\hbox{otherwise},
\end{array}\right.
\end{eqnarray}
where $\varepsilon_{\textup{eff}}$ is the effective dielectric permittivity inside the ion and is thought to be related with the ionic specificity, e.g., ionic polarizability. The ions can also be treated as conducting spheres with adjustable hydrated radius. This treatment has been shown to perform well in matching the simulation and experimental results and explaining the Hofmeister effect for ions near dielectric interfaces \cite{Levin:PRL1:09,Levin:PRL2:09,WW:PRL:14}. However, we will not investigate polarizable effects in this work and assume a constant interior dielectric constant $\varepsilon_{\textup{eff}}=\varepsilon_0$.
%where $\varepsilon_0$ is the effective dielectric permittivity inside the ion and we takes the vacuum dielectric constant %in this work,  $a_i$ is the radius of $i$th ion.
At the point charge assumption $a_i\rightarrow 0$, Eq. \eqref{ugfs} is divergent for a space-dependent $\varepsilon$, thus this modification is essential to ensure a finite self energy $u_i$. It can be emphasized that ionic specific effects can be accounted for by defining a species-dependent $\varepsilon_0$ in Eq. \eqref{eq:ei}.

The large potentials usually appear near the surfaces with strong surface charges, the treatment with steric entropy is useful to control numerical stability and performs well to constrain the concentration of counterions in a physically significant regime. For the case of asymmetric ionic sizes, we take the lattice volume $\nu$ as the volume of a counterion, considering the ion distribution is far below the saturation density and not sensitive to the size of the coions. On the other hand, we do introduce the effect of asymmetry in ion size when calculating the self energy with Eq. \eqref{gifs}. We find this consideration is necessary when the ion-ion correlation between coions is also strong, e.g., for 2:2 electrolytes.

\subsection{Nondimensionalization} \label{sec:nondim}

Let $L$ be a length scale to characterize the geometric length of interfaces, $\ell_B=\beta e^2/(4\pi \varepsilon_W)$ be the Bjerrum length in water solvent, and $\ell_D=1/\sqrt{4\pi\ell_B\sum_i z_i^2 c_{i0}}$ be the Debye screening length.
Following dimensionless parameters and variables in \cite{XML:PRE:2014}, we define
$\widetilde{\textbf{r}}=\textbf{r}/L$,  $\widetilde{c}_{i0}= c_{i0}/c_{+0}$,  $\widetilde{\varepsilon}=\varepsilon/\varepsilon_W$, $\widetilde{\Phi}=\beta e \Phi$, $\widetilde{G}_i= \beta e^2 G_i$, and $\widetilde{G}_0=\beta e^2 G_0$.   We define the dimensionless variables for the surface charge density $\widetilde{\sigma}= \sigma L^2/e$ and the dipole moment $\widetilde{p}_0=p_0/(eL)$ for the use in next sections. We drop
%the
tildes of all new variables and have the dimensionless modified PB and DH equations as the following,
\begin{eqnarray}
&&-2I_0\epsilon^2\nabla\cdot \varepsilon\nabla\Phi = \sum_i z_i c_{i0}e^{-U_i-S}, \label{eq:dlSEMPB}\\
&&U_i=z_i\Phi+\frac{1}{2}z_i^2u_i, \label{eq:Ui}\\
&&S=\log\big[1+\sum_j \nu c_{j0}\left( e^{-U_j}-1 \right)\big], \label{eq:S}\\
&&u_i = \lim_{\mathbf{r}'\rightarrow\mathbf{r}}\left[G_i(\mathbf{r},\mathbf{r}')-G_0(\mathbf{r},\mathbf{r}')\right], \label{eq:uGG0}\\
&&-\nabla \cdot \varepsilon_i  \nabla G_i + I_i /(I_0\epsilon^2) G_i = 4\pi q \delta(\mathbf{r}-\mathbf{r}'), \label{eq:dlDH}
\end{eqnarray}
where
\begin{eqnarray}
&&G_0=\frac{q}{\varepsilon_\textup{eff}|\mathbf{r}-\mathbf{r}'|}, \label{eq:dlG0}\\
&& \varepsilon_i(\mathbf{r},\mathbf{r}') =
\left\{ \begin{array}{ll}
                \varepsilon_\textup{eff},~~~~~~~~~~ |\mathbf{r}-\mathbf{r}'|<a_i,\\
                \varepsilon(\mathbf{r}),~~~~~~~~~ \hbox{otherwise},
\end{array}\right. \label{eq:eps}\\
&&    I_i(\mathbf{r},\mathbf{r}') = \left\{ \begin{array}{ll}
                0,~~~~~~~~~~~~~|\mathbf{r}-\mathbf{r}'|<a_i,\\
                \frac{1}{2}\sum_i z_i^2c_i,~~\hbox{otherwise},
\end{array}\right.
\end{eqnarray}
and $\epsilon = \ell_D/L$ and $q=\ell_B/L$ are two dimensionless parameters, $I_0=\sum_iz_i^2c_{i0}/(2c_{+0})=z_+(z_++1)/2$ is the scaled far field ionic strength, which is one for 1:1 electrolytes, and 3 for 2:1 electrolytes. We shall note that after the nondimensionalization the dielectric permittivities $\varepsilon_\textup{eff}=1/80$ and $\varepsilon(\mathbf{r}) \rightarrow 1,$
since they are divided by the water dielectric constant. It is not difficult to see that $(2I_0\epsilon^2 \varepsilon)$ denotes the effective dielectric function, and $q$ represents the effective charge of the test ion and hence the strength of its self energy.
Eqs. \eqref{eq:dlSEMPB}-\eqref{eq:dlDH} comprises the SCF model for variable media. This system of self-consistent equations is general for different electrolytes and arbitrary surface geometries. The solution of the DH equation is challenging, not only because of its high dimensions but also because it is a multiscale problem, i.e., the sizes of nanoparticles and mobile ions are of two different spatial scales.

\subsection{The self-energy approximation}

To approximate $u_i$, we denote an inverse Debye length function by $\kappa(\mathbf{r})=\sqrt{\sum_iz_i^2c_i/(2I_0\epsilon^2)}$,
and set $q=1$ without loss of generality,
then the DH equation is written as,
\begin{equation}
-\nabla \cdot \varepsilon_i \nabla G_i +\kappa_i^2 G_i = 4\pi\delta(\mathbf{r}-\mathbf{r}'),
\end{equation}
where similar to $\varepsilon_i$ defined by Eq. \eqref{eq:eps}, $\kappa_i(\mathbf{r},\mathbf{r}')$ is six dimensional and locally depend on the location of the test ion,
\begin{equation}
    \kappa_i(\mathbf{r},\mathbf{r}') = \left\{ \begin{array}{ll}
                0,~~~~~~~~~~~|\mathbf{r}-\mathbf{r}'|<a_i,\\
                \kappa(\mathbf{r}),~~~~~~\hbox{otherwise}.
\end{array}\right.
\end{equation}
We recall the dimensionless $\varepsilon$ is unitary in the bulk solvent region.

To resolve the multiscale problem in the approximation of the self energy, we decompose the solution for $u_i$ into two parts by
\begin{eqnarray}
u_i &=& u_1 + u_2
\nonumber\\
 &=& \lim_{\mathbf{r}'\rightarrow\mathbf{r}}(G_i-G_i')+\lim_{\mathbf{r}'\rightarrow\mathbf{r}}(G_i'-G_0), \label{eq:u}
\end{eqnarray}
such that $G_i'$ satisfies,
\begin{equation}
-\nabla \cdot \varepsilon'_i\nabla G_i' = 4\pi\delta(\mathbf{r}-\mathbf{r}'),
\end{equation}
where
\begin{equation}
    \varepsilon'_i(\mathbf{r},\mathbf{r}') = \left\{ \begin{array}{ll}
                \varepsilon_\textup{eff},~~~~~~~~~|\mathbf{r}-\mathbf{r}'|<a_i,\\
                \varepsilon (\mathbf{r}'),~~~~~~\hbox{otherwise}.
\end{array}\right.
\end{equation}
We shall see that $\varepsilon'_i$ is piecewise constant, which is uniform outside the ionic cavity. Then
the second part of the self energy, $u_2$, can be evaluated locally and analytically, giving us a form of the Born energy \cite{Born:ZP:1920},
\begin{equation}
    u_2  =  \frac{1}{a_i}\left[\frac{1}{\varepsilon(\mathbf{r})} - \frac{1}{\varepsilon_\textup{eff}}\right],
\end{equation}
which characterizes the mutual effect of the ionic size and variable permittivity.

In the first component of the self energy \eqref{eq:u}, the solution of Green's function depends on the boundary, global ionic concentrations and dielectric permittivity, and $u_1$ has no explicit expression. We consider the small $a_i$ asymptotic and that $\varepsilon'_i\approx \varepsilon_i$ in the locality of $\mathbf{r}'$. At the limit $a_i\rightarrow 0$, the self Green's function $\widetilde{u}_1=\lim_{\mathbf{r}'\rightarrow\mathbf{r}}(\widetilde{G}-\widetilde{G}')$,
where
\begin{equation}
-\nabla \cdot \varepsilon(\mathbf{r}) \nabla \widetilde{G} +\kappa^2(\mathbf{r}) \widetilde{G} = 4\pi\delta(\mathbf{r}-\mathbf{r}'), \label{tgreen}
\end{equation}
and $\widetilde{G}'=1/\varepsilon(\mathbf{r}')|\mathbf{r}-\mathbf{r}'|$.
Here, $\varepsilon(\mathbf{r})$ and $\kappa(\mathbf{r})$ do not depend on the site of the test ion, and we then remove the difficulty of solving two space scales. We shall use the far field approximation to include the excluded-volume effect in the definition of $\kappa_i(\mathbf{r},\mathbf{r}')$, which yields the expression for $u_1$,
\begin{equation}
	u_1 \approx \frac{\widetilde{u}_1}{1-\varepsilon(\mathbf{r})\widetilde{u}_1 a_i}. \label{eq:approx_u1}
\end{equation}
The approximate expression \eqref{eq:approx_u1} is accurate in the sense that the function $\kappa(\mathbf{r})$ is not rapidly varying or the ion size $a_i$ is small.

The only expensive part is then how to solve Eq. \eqref{tgreen} to obtain $\widetilde{u}_1 = \displaystyle \lim_{\mathbf{r}'\rightarrow\mathbf{r}}(\widetilde{G}-\widetilde{G}')$. Let
\begin{equation}
\mathcal{H}=\sqrt{\varepsilon(\mathbf{r})\varepsilon(\mathbf{r}')}  \widetilde{G}.
\end{equation}
Since $\varepsilon(\mathbf{r})$ and $\kappa(\mathbf{r})$ are both smooth functions,
we transform the Green's function equation into the equation for $\mathcal{H}$,
\begin{equation}
-\nabla^2 \mathcal{H}+v^2 \mathcal{H}= 4\pi\delta(\mathbf{r}-\mathbf{r}'),\label{eq:H}
\end{equation}
where, after using the property $g(\mathbf{r})\delta(\mathbf{r}-\mathbf{r}')=g(\mathbf{r}')\delta(\mathbf{r}-\mathbf{r}')$ for any smooth $g$, we can write,
\begin{equation}
v(\mathbf{r})=\left[\frac{\sqrt{\varepsilon(\mathbf{r})}\nabla^2\sqrt{\varepsilon(\mathbf{r})}+\kappa^2(\mathbf{r})}{ \varepsilon(\mathbf{r}) } \right]^{1/2}.\end{equation}
For the free-space Green's function, we introduce $\mathcal{H}_0=\varepsilon(\mathbf{r}')\widetilde{G}'=1/|\mathbf{r}-\mathbf{r}'|$.
Subtracting Eq. \eqref{eq:H} and $\mathcal{H}_0$ into Eq. \eqref{eq:H} gives,
\begin{equation}
 \nabla^2 (\mathcal{H}-\mathcal{H}_0) =  v^2 \mathcal{H}.
\end{equation}
The solution of this equation can be represented by a Born series through the integral equation,
\begin{equation}
\mathcal{H}-\mathcal{H}_0  = -\frac{1}{4\pi}\int \frac{v^2(\mathbf{r}'')\mathcal{H}(\mathbf{r}'',\mathbf{r}')}{|\mathbf{r}-\mathbf{r}''|}  d\mathbf{r}''.
\end{equation}
To solve it, initially we could take the leading asymptotic for $\mathcal{H}$,
\begin{equation}
\mathcal{H}^{(1)}(\mathbf{r},\mathbf{r}')=\frac{\exp[v(\mathbf{r})r_{12}]}
{r_{12}},
\end{equation}
where $r_{12}$ is the distance between $\mathbf{r}$ and $\mathbf{r}'$,
then a series solution is iteratively defined by,
\begin{equation}
\mathcal{H}^{(n)}-\mathcal{H}_0  = -\frac{1}{4\pi}\int \frac{v^2(\mathbf{r}'')\mathcal{H}^{(n-1)}(\mathbf{r}'',\mathbf{r}')}{|\mathbf{r}-\mathbf{r}''|} d\mathbf{r}''. \label{eq:BornSeries}
\end{equation}
In principle, the solution is obtained upon the convergence of the iteration. Each step includes a multiplication of two dense matrices, which is too expensive to implement. However, $\mathcal{H}^{(1)}$ is actually pretty good asymptotic approximation to the Green's function, and we could expect that $\mathcal{H}^{(2)}$ may include enough information to understand the Green's function.
At this point, the expression for $\widetilde{u}_1$ is,
\begin{eqnarray}
&\widetilde{u}_1 & = \frac{1}{\varepsilon(\mathbf{r})} \lim_{\mathbf{r}'\rightarrow\mathbf{r}}[\mathcal{H}^{(2)}(\mathbf{r},\mathbf{r}')-\mathcal{H}_0(\mathbf{r},\mathbf{r}')] \nonumber \\
&&= -\frac{1}{4\pi\varepsilon(\mathbf{r})}\int \frac{v^2(\mathbf{r}'')\mathcal{H}^{(1)}(\mathbf{r}'',\mathbf{r})}{|\mathbf{r}-\mathbf{r}''|} d\mathbf{r}''. \label{tildeu1}
\end{eqnarray}
Finally, we find an approximate solution for the self energy,
\begin{equation}
	u_i = \frac{\widetilde{u}_1}{1-\varepsilon(\mathbf{r})\widetilde{u}_1 a_i}+\frac{1}{a_i}\left[\frac{1}{\varepsilon(\mathbf{r})} - \frac{1}{\varepsilon_\textup{eff}}\right],~i=\pm, \label{eq:uasym}
\end{equation}
where $\widetilde{u}_1$ is tge integral expression \eqref{tildeu1}.
The global dielectric and ionic strength variations due to dielectric interfaces are included in $v$.
For general $q$, Eq. \eqref{eq:uasym} multiplied by $q$ is the self energy of the ion.

For weak coupling cases, e.g., dilute electrolytes with monovalent ions, the Born energy (the second term in Eq. \eqref{eq:uasym}) dominates the self energy. This term is a local property which only depends on the local dielectric constant and ionic radius. This energy has been incorporated into the classical mean-field PB theory by Paunov \textit{et al.} \cite{PDKBM:JCIS:96}. This modified PB theory captures some important effects due to inhomgeneous dielectric media in dilute electrolytes. The modified equation can be solved simply by the iterative numerical methods. However, for strong coupling cases as discussed in the present work, the nonlocal correlation effect (the first term in Eq. \eqref{eq:uasym}) is of great importance and non-ignorable. As is shown, the correlation energy corresponds to the solution of the self Green's function. In order to approximate the sophisticated procedure of solving the generalized DH equation \eqref{eq:dlDH}, we make use of the concept of Born series \eqref{eq:BornSeries} and arrive at the integral formula as \eqref{tildeu1}, which corresponds to a matrix-vector multiplication in discretization form, and thus is achievable in practical use.

It should be remarked that the convergence of the Born series depends on the smoothness of the dielectric permittivity. The accuracy may be low when a large gradient is present in the dielectric profile. When the interface is regular such as planes or spheres, this can be tackled by taking the integral region as the solvent region and introducing a WKB-type approximation by using the image charge method; see \cite{XML:PRE:2014} for planar interfaces. For a general geometry, an efficient approximation remains an open question.

\subsection{Field-dependent dielectric permittivity}

In this section, we introduce the dipolar model for the dielectric permittivity. The solvent molecules of the solution generally contain permanent dipoles. When an external electric field is applied, the dipoles are oriented against the field and the solvent is polarized. As aforementioned in the Introduction, the dependence of polarization and dielectric permittivity on the electric field strength
was
systematically discussed by Booth \cite{B:JCP:1951, B:JCP:1955} in the 1950s. In the dimensional form,
the Booth's relation is expressed by the formula,
\begin{equation}
	\varepsilon(E) =  \varepsilon_0 + \frac{p_0n_d(\beta p_0E)\mathscr{L}(\beta p_0E)}{E} , \label{eq:booth}
\end{equation}
with the field strength $E=|\nabla\Phi|$, the Langevin function $\mathscr{L}(x)=\coth x - 1/x$ and the dipole density $n_d(x) = c_d\sinh x/x$, where $p_0$ and $c_d$ are the permanent dipole moment of a solvent molecule and its number density. Considering the case of water solvent, we have the limit in bulk electrolytes $\varepsilon(E \rightarrow 0)=\varepsilon_0+\beta p_0^2c_d/3=\varepsilon_W$, which gives $\beta p_0^2 c_d = 3(\varepsilon_W-\varepsilon_0)$.
In Eq. \eqref{eq:booth}, we have neglected the finite size effect of solvent molecules by assuming them as point dipoles. This field-dependent dielectric effect was involved into the classical PB model by Paunov \textit{et al.} \cite{PDKBM:JCIS:96} and was named later as dipolar PB (DPB)  equation by Abrashkin \textit{et al.} \cite{AAO:PRL:2007}.
Usually, it is reasonable to neglect the small change in the dipolar density for water solvent,
i.e., assume $n_d(\beta p_0E)=c_d$, which yields the Langevin PB equation \cite{FO:PCCP:11}. Interestingly, this neglect repairs the divergence of the dipole density, and thus the Langevin equation gives more reasonable prediction for the dielectric function than the diploar model \cite{Frydel:review:2014}. 

We then follow the simplified version for ignoring the density fluctuation of water, and include the self-energy contribution due to the dielectric inhomogeneity and the electrostatic correlation. In dimensionless form, the Langevin-SCF equation (SCF-L) reads,
\begin{equation}
 -2I_0\epsilon^2\nabla \cdot \varepsilon(|\nabla\Phi|)\nabla\Phi = \sum_i z_i c_{i0}e^{-U_i-S}, \label{eq:scfl}
\end{equation}
where $U_i$ and $S$ are expressions described in Eqs. \eqref{eq:Ui} and \eqref{eq:S} and
\begin{equation}
\varepsilon(E) = \varepsilon_0+3(1-\varepsilon_0) \mathscr{L}(p_0 E)/(p_0 E).
\end{equation}
We have $\varepsilon_0=1/80$ and $\varepsilon(E\rightarrow 0)=1$.

Besides the above field-dependent or Booth's model for dielectric permittivity, there have been many discussions for other effects to local variations of the dielectric permittivity.
With the simple dipole moment model, one can see that ions in electrolytes undergo solvation shells and prevent the surrounding water molecules from being oriented against the external field. The dielectric permittivity of the solution will be decreased when ions are added to it, and one can obtain equations for the salt-concentration-dependent dielectric permittivity for electrolytes \cite{HRC:JCP:48,BR:NY:98,KD:JCP:09,BAP:JCP:11,LWZ:CMS:14,Renou:JPCB:14}. Furthermore, the ionic polarization also plays a role in the dielectric function because ions act
as induced dipole moments in an external field \cite{Frydel:JCP:11}. When these effects  are considered, the potential of mean force $U_i$ in the SCF model has to be modified again. However, we will limit the discussion to Eq. \eqref{eq:scfl} and save them as future work if these effects have to be included to describe unexplored physics.

\section{Electrolytes around a macroion}

\begin{figure}[h]
\includegraphics[scale=0.33]{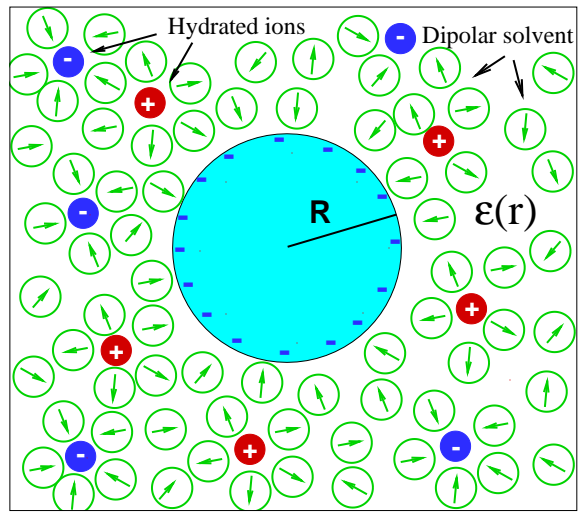}
\caption{Schematic representation of a charged nanoparticle immersed in an electrolyte. Cations and anions are represented by solid red and blue balls, respectively. The surface charges are negative. Dipolar solvent molecules (empty green balls) near the surface are ordered under the influence of the strong electric field, leading to nonuniform dielectric permittivity. }
\label{fig:schem}
\end{figure}

Now we consider a special case of a charged nanoparticle with radius $R$ surrounded by an electrolyte (see Fig. \ref{fig:schem}). The surface charge is distributed uniformly on the surface with density $\sigma$ to make the potential field spherically symmetric. Thus, the potential field and ionic concentration densities
are spherically symmetric in spherical coordinates $(r,\theta,\phi)$.
We have the Neumann boundary condition for electric potential on the surface
\begin{equation}
	-\varepsilon(r) \left.\frac{\partial\Phi}{\partial r} \right|_{r=R}= 4\pi q \sigma.
\end{equation}
Note here we have used the dimensionless surface charge density and $q=\ell_B/L$, and the dielectric permittivity is also dimensionless which is relative to the water dielectric permittivity.
The origin of the sphere is set to be $r=0$ in the spherical coordinates. Clearly, we have the fact that the $i$th ion cannot penetrate the region $r<R+a_i$ due to the hard-wall repulsion to mobile ions. However, in order to maintain the continuous density profiles, we simplify this boundary effect for asymmetric electrolytes and assume the modified boundary is located at $r = R + a$ with $a = \max a_i$. Thus, the modified boundary condition is
\begin{equation}
	-\varepsilon(r) \left.\frac{\partial\Phi}{\partial r} \right|_{r=R+a}= 4\pi q \sigma\frac{R^2}{(R+a)^2}.
\end{equation}
The far field conditions for electric potential is the decaying condition, $\Phi(r\rightarrow+\infty)=0$.

In this special spherical geometry, the dimensions in the self energy $u_i$ can be reduced,
where $\widetilde{u}_1$  in Eq. \eqref{eq:uasym} is simplified as,
\begin{eqnarray}
&\widetilde{u}_1&= \frac{1}{\varepsilon}\lim_{\mathbf{r}'\rightarrow\mathbf{r}}[\mathcal{H}^{(2)}(\mathbf{r},\mathbf{r}')-\mathcal{H}_0(\mathbf{r},\mathbf{r}')] \nonumber\\
%& &= -\frac{1}{4\pi\varepsilon}\int v^2(r'')\frac{e^{-v_0(r)\sqrt{{r''} ^2+r-2rr''\cos\theta}}}{{r''}^2+r-2rr''\cos\theta}{r''}^2\sin\theta dr''d\theta d\phi \nonumber\\
&&= -\frac{1}{2\varepsilon}\int v^2(r'')\frac{e^{-v_0(r)\sqrt{{r''} ^2+r-2rr''\cos\theta}}}{{r''}^2+r-2rr''\cos\theta}{r''}^2\sin\theta dr''d\theta \nonumber\\
&&= -\frac{1}{2\varepsilon}\int_0^{+\infty} v^2(r'')\frac{r''}{r} dr'' \int^{v^2(r'')|r+r''|}_{v^2(r'')|r-r''|} \frac{e^{-z}}{z}dz. \label{eq:int}
\end{eqnarray}
The exponential integral is a special function and can be pre-computed.
The computational cost of evaluating $\delta \mathcal{H}(\mathbf{r}\rightarrow\mathbf{r}')$ for each point $\mathbf{r}'$ becomes only a one-dimensional integral for $r''$.

We propose a self-consistent iterative algorithm to solve the SCF and SCF-L models. The algorithm is composed of two alternating steps: (1) For given $c_i^{(k)}$ and $\Phi^{(k)}$, we update $u_i^{(k)}$; and (2) for given $u_i^{(k)}$, we solve the modified PB equation for $\Phi^{(k+1)}$ subject to given boundary conditions.
The iterative scheme for time steps is as follows,
\begin{eqnarray}
&& u_i^{(k)} = \frac{\widetilde{u}_1^{(k)}}{1-\varepsilon^{(k)} \widetilde{u}_1^{(k)} a_i}+\frac{1}{a_i}\left[\frac{1}{\varepsilon^{(k)} } - \frac{1}{\varepsilon_0}\right],\\
&&\widetilde{u}_1^{(k)} = - \int \frac{[v^{(k)}(\mathbf{r}'')]^2\exp[v^{(k)}(\mathbf{r})|\mathbf{r}-\mathbf{r}''|]}{4\pi\varepsilon^{(k)}(\mathbf{r})|\mathbf{r}-\mathbf{r}''|^2} d\mathbf{r}'',\\
&&-2I_0\epsilon^2\nabla\cdot \varepsilon^{(k+1)}\nabla\Phi^{(k+1)} = \sum_i z_i c_{i0} e^{-U_i^{(k+1)}-S^{(k+1)}},     \nonumber\\
\label{eq:pb}\\
&&U_i^{(k+1)}=z_i\Phi^{(k+1)}+\frac{1}{2}z_i^2q u_i^{(k)},\\
&&S^{(k+1)}=\log\big[1+\sum_j \nu c_{j0}\big( e^{-U_j^{(k+1)}}-1 \big)\big],
\end{eqnarray}
for $k=0,1,2,...,K$ and the iterative steps are performed until the convergence criteria $|u_i^{(k+1)}-u_i^{(k)}|<\delta$ is arrived, where $\delta$ is a small tolerant value. Here, we shall note that $\varepsilon$ and $v$ are functions of $\Phi$ and $c_i$.
This numerical scheme follows a recent work \cite{XuMaggs:JCP:14} for solvents of uniform dielectric function where the self-Green's function in the DH equation can be efficiently solved by a method of the selected inversion. Differently, in this paper a numerical integration is used for the self energy, thus we can work on more general geometries.

The sub-level of the iterative scheme is embodied in the solution of the modified PB equation \eqref{eq:pb},  which shows high nonlinearity. We could in general use the iterative scheme at the $(k)$th loop as
\begin{equation}
-2I_0\epsilon^2\nabla\cdot \varepsilon^{[l]}\nabla\Phi^{[l+1]}-\gamma^{[l]}\Phi^{[l+1]} = \sum_i z_i c_{i}^{[l]} -\gamma^{[l]}\Phi^{[l]},  \label{mpb:space}
\end{equation}
where $c_{i}^{[l]}$ means the potential in $U_i$ and $S$ uses the value from the $l$th step,
and the relaxation function $\gamma(\textbf{r})$ is defined as
\begin{equation}
	\gamma(\textbf{r}) = |\sum_iz_ic_i| / (|\Phi|+\delta_0),
\end{equation}
with $\delta_0=10^{-8}$.
Here we use the index $[\cdot]$ to represent the sub-iteration in the $(k)$th loop of the self-consistent iteration.
This iteration stops when the convergence criteria $|\Phi^{[l+1]}-\Phi^{[l]}|<\delta$ is arrived.

For the physical setting with one macroion, the mean potential is spherically symmetric, hence the space discretization for Eq. \eqref{mpb:space} can be done by the central difference scheme for one-dimensional equation along the radial direction. The composite trapezoidal rule is used to approximate the integral in computing Eq. \eqref{eq:int} for $\tilde{u}_1$.

We have not mentioned the direct numerical method for solving the self Green's function. Solving the diagonal elements of lattice Green's function from the generalized DH equation can be done by the use of selected inversion \cite{LYL+:SJoSC:2011,LYM+:ATMS:2011}, which has been shown to reduce the computational cost a lot with point charge limit \cite{XuMaggs:JCP:14,XML:PRE:2014}. However, direct solvers of problems with inhomogeneous dielectric constant and size effect still remain
challenging since it is a two-scale problem and we have to use a mesh size much smaller than an ion radius.

\section{Results}

In this section, we present numerical results for different systems of binary electrolytes surrounding a spherical macroion. We study the position-dependent dielectric constant $\varepsilon(\mathbf{r})$, compared with results from
MC simulations. We then perform calculations
for the SCF model with the
field-dependent dielectric function. We use 1:1 and 2:1 salts, where anions are coions and always monovalent. Divalent counterions are used for strongly correlated systems, for which a correct treatment of electrostatic self energy is essential. Without specific statement, ions have radius of $a=0.225~\textup{nm}$, the simulations are performed at room temperature and the relative dielectric constant
is $80$ for the bulk water, so the Bjerrum length $\ell_B=7.14~\textup{nm}$.
The dimensional dipole moment is chosen as $p_0=4.8~ \textup{D}$ \citep{AAO:PRL:2007} for keeping the bulk water density $c_d=55~\textup{M}$. Surface charge density is varied within $0< -\sigma < 3~\textup{e/nm}^2$, which describes most of the environments in biological or physical systems.

\subsection{Space-dependent dielectric function}

In the following two groups of simulations, we consider simulated systems with given space-dependent dielectric functions. We take the radius of macroion as
$R=2~\textup{nm}$, and the grid size for the SCF approximation as
$3.75\times 10^{-3}~\textup{nm}.$

\begin{figure}[t!]
\includegraphics[scale=.33]{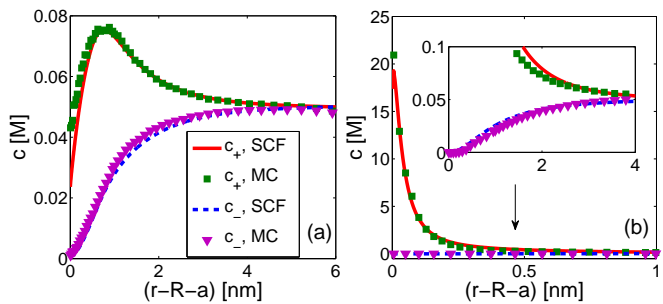}
\caption{Counterion and coion concentration profiles as a function of the distance to the surface in 1:1 symmetric electrolytes with bulk concentration $c_\pm = 50$ mM and surface charge $\sigma = -0.2~ \textup{e/nm}^2$ (a) and  $\sigma = -1.6 ~\textup{e/nm}^2$ (b). The ramp shape of dielectric profile is given by Eq. \eqref{eq:rampdc}. Both SCF and MC simulations are presented. }
\label{fig:11c50}
\end{figure}

{\it Electrolyte with 1:1 salts.}
We first consider that the macroion is surrounded by an electrolyte of  $50~\textup{mM}$ 1:1 salts. A ramp shape of radial function is used to describe the dielectric permittivity, given by the following formula,
\begin{equation}
	\varepsilon(r) = \varepsilon_W+ \frac{\varepsilon_C-\varepsilon_W}{1+\exp[(r-2.45)/\lambda]}, \label{eq:rampdc}
\end{equation}
where the unit for the spatial length is $\textup{nm}$, and $\lambda=0.2~\textup{nm}$ is the length scale for the region of varying dielectric. We shall see the dielectric constant is $\varepsilon_C$ in the deep of the macroion, and $\varepsilon_W$ in the bulk water. Experimental measurements and computational studies have indicated that the dielectric permittivities of various ionic fluids near different interfaces can be fitted by the exponential function \cite{TCS:PRE:01,PCZ:JCP:87}. We take $\varepsilon_C=10\varepsilon_0$ and $\varepsilon_W=80\varepsilon_0.$ In Fig. \ref{fig:11c50}, the counterion and coion density distributions along the radial direction are plotted by numerically solving the SCF equations and particle-based MC simulations.  Two sets of surface charge densities, $\sigma=-0.2$ and $-1.6~ \textup{e/nm}^2$, are taken to check the accuracy of the SCF computation. To run the MC simulation, the harmonic interpolation method (HIM) \cite{FXH:JCP:14} is applied to calculate ion-ion pairwise energies with well treated boundary conditions for the Hamiltonian. The simulation volume is a concentric cell of radius $R_c=8.629~\textup{nm}$.

For a surface of low surface charge, the electrostatic attraction between the macroion and counterions is weak, while the Born solvation energy increases when an ion approaches to the surface. It can be seen from \ref{fig:11c50}(a) that the counterion is depleted near the surface, leading to an obvious nonmonotonic counterion concentration. For
the strong surface charge, however, the attraction from the macroion dominates, clear depletion effect is not observed from \ref{fig:11c50}(b). %The SCF and MC results in both weak and strong coupling cases are shown %in a good agreement with each other.
The SCF results in both weak and strong coupling cases are shown in good agreement with MC simulations.

{\it Electrolyte with 2:1 salts.}
In the second group of systems, we consider a sharp jump of dielectric permittivity across the macroion surface with $\varepsilon_C=2\varepsilon_0$ and $\varepsilon_W=80\varepsilon_0$ inside and outside it, and an electrolyte of $50~\textup{mM}$ 2:1 salts which will show strong correlation effects for high surface charges. The dielectric-boundary effect is of interest in community \cite{HL:SM:2008,Messina:JPCM:2009,WM:JPCB:2010,GXX:JCP:2012,JSC:JCP:13,WangRui:JCP:13,ZD:PNAS:13}.
For example, the theoretical treatment of dielectric-boundary effects for both weak and strong coupling cases was also developed \cite{Bakhshandeh:PRL:2011}. It is then of interest to study the interplay between dielectric-boundary (image charge) and correlation effects in the SCF framework, verified with MC simulations. 
Two surface charge densities $\sigma=-0.2$ and $-2~\textup{e/nm}^2$ are taken. The radius of the simulation cell for MC calculations is $10~\textup{nm}$. For this setup, the image charge method \cite{CDJ:JCP:2007,GX:PRE:2011} is used to obtain accurate MC results.
As aforementioned, the Born series for the self energy has slow convergence for the sharp dielectric variation and Eq. \eqref{eq:int} fails when it integrates over the whole domain. In order to make the Born series work, we take the lower limit of integral Eq. \eqref{eq:int} as $R+a$ for $\widetilde{u}_1$, and follow similar technique for planar interfaces \cite{XML:PRE:2014} to introduce the dielectric effect from the macroion by a WKB type approximation using image charges for an updated self energy,
\begin{equation}
\widetilde{u}_1'= \widetilde{u}_1 + \sum_{\ell=1}^L \frac{z_\ell \exp(\widetilde{u}_1 D_\ell)}{\varepsilon_W D_\ell}, \label{wkb}\end{equation}
where index $\ell$ means the $\ell$th image charge,
$z_\ell$ is its strength due to a unit source charge, and $D_\ell$ is its distance to the source charge; for these parameters see Eqs. (11)-(12) in Ref. \cite{GX:PRE:2011}. Both SCF and MC simulations use three image charges.

\begin{figure}[t!]
\includegraphics[scale=.332]{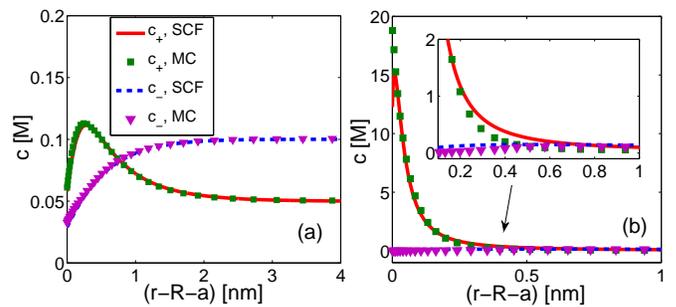}
\caption{Counterion and coion concentration profiles as a function of the distance to the surface in 2:1 symmetry electrolytes with bulk concentration $c_+ = 50$ mM and surface charge $\sigma = -0.2~ \textup{e/nm}^2$ (a) and  $\sigma = -2 ~\textup{e/nm}^2$ (b). A sharp dielectric jump is across the colloidal surface. Both SCF and MC simulations are presented.}
\label{fig:21c50}
\end{figure}

The results are presented in Fig. \ref{fig:21c50} for both weak and strong surface charge densities. From the left panel, we shall see that the WKB type approximation is very accurate for the dielectric effect, and the SCF and MC results are in a perfect agreement. For the weak-coupling system (Fig. \ref{fig:21c50}(a)), the image charges repel mobile ions and a depletion zone is formed near the surface. For the strong-coupling system (Fig. \ref{fig:21c50}(b)), they are screened by the surface charge since the image charges are dipole effects.
The MC simulation does not predict a depletion zone, while the SCF result has a short interval for the depletion due to the steric effect applied to the double-layer ions in the vicinity of the surface. Both results show much higher counterion density near the surface than the bulk region with a factor of $\sim300$ times ($15~\textup{M}$ vs. $50~\textup{mM}$). Both %the
coion distribution profiles are slightly nonmonotonic, indicating a possibility of overscreening inside the double layer. These nonmonotonic curves are important signals for the correlation-induced charge inversion widely studied in literature \cite{BZH+:PRL:2004,PBV+:PRL:2006,Messina:JCP:2002,Tanaka:PRE:2003,DL:JCP:2006,LH:EPJE:2008,GXX:JCP:2012}. The results shown in the two panels demonstrate that the SCF model can accurately capture the dielectric-boundary effect and the ion-ion electrostatic correlation.

\subsection{Field-dependent dielectric function}

We have seen that the SCF theory can accurately capture the effects of electrostatic correlation, dielectric boundary and varying dielectric permittivity by a comparison with particle-base MC simulations with a fixed space-dependent dielectric function. In this section, we study equilibrium properties for systems of field-dependent dielectric function using the Langevin-SCF model. We consider both 1:1 and 2:1 salts with a fixed salt density $100~\textup{mM}$ and set the radius of the macroion, $R=4~\textup{nm}$. The dielectric-boundary effect is ignored by assuming the dielectric constant for $r<R+a$ is uniform and equal to $\varepsilon(R+a)$, which greatly improves the convergence of the Born series. For the numerical approximation, we take the grid size as
$2.5\times 10^{-3}~\textup{nm}$.

\subsubsection{Effect of the ion correlation}

\begin{figure}[t!]
\includegraphics[scale=.40]{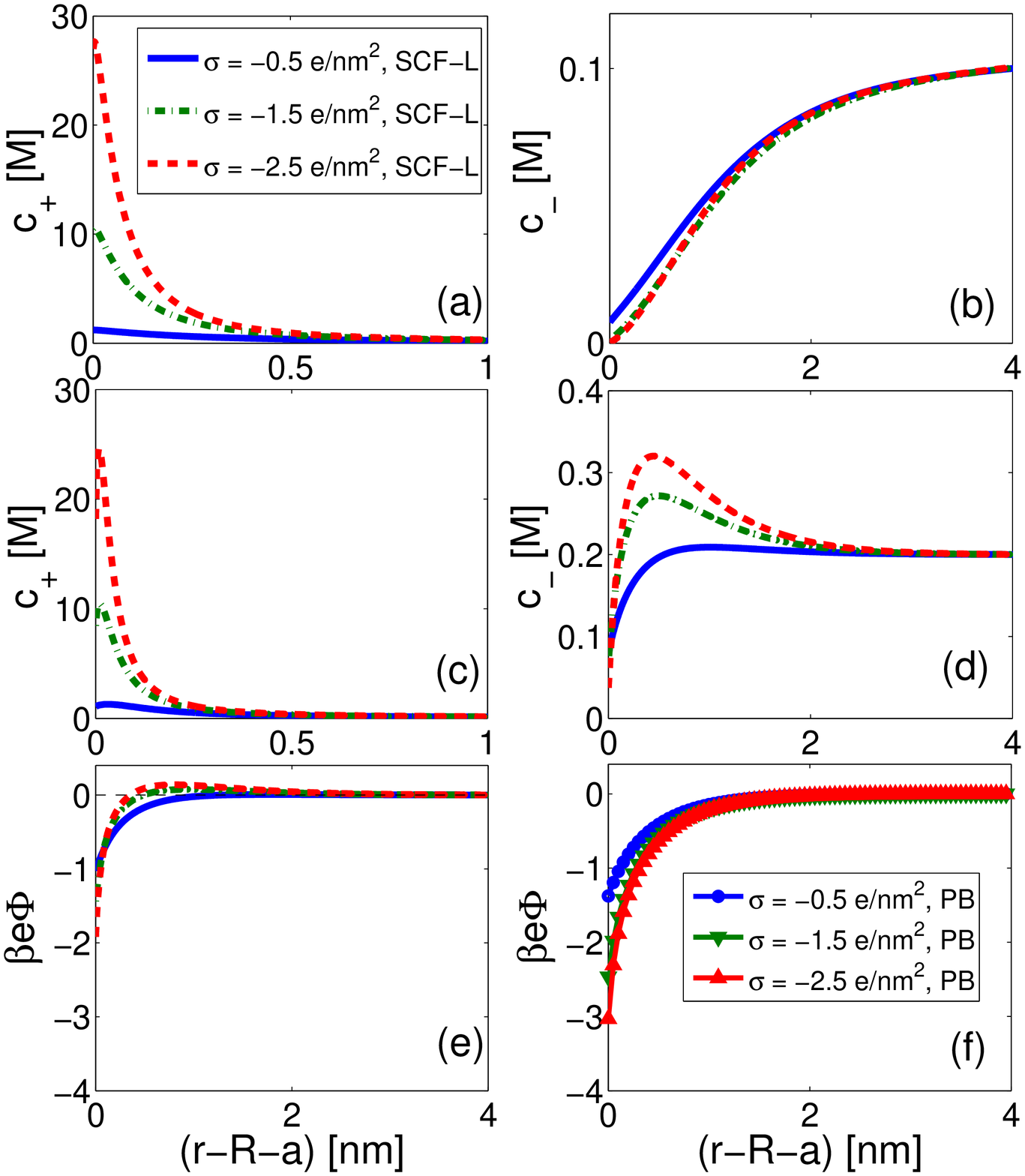}
\caption{
Ion concentrations and electric potentials for SCF calculations with field-dependent dielectric function (SCF-L). (a)(b) Ion concentrations for 1:1 salts; (c)(d) Ion concentrations for 2:1 concentration; (e)(f) Electric potential profiles for 2:1 salts by SCF-L and classical PB models.
} \label{fig:11symphiden}
\end{figure}

In Fig. \ref{fig:11symphiden}, numerical results of ionic concentrations and electric potentials for three different surface charges $\sigma=-0.5, -1.5$ and $-2.5~\textup{e/nm}^2$ are given. Correspondingly, the coupling parameters for 1:1 salts are $\Xi=1.6$, 4.8 and 8.0, and those for 2:1 salts are $\Xi=12.8$, 38.4 and 64.1. The upper and middle panels show the cation and anion density distributions, (a)(b) for 1:1 salts, and (c)(d) for 2:1 salts. The bottom panels, (e)(f), are the electric potentials for 2:1 salts from both the SCF-L and classical PB predictions.

For monovalent ions, since the electric field only slightly changes the nearby permittivity at the surface, the Born-energy repulsion to mobile ions is weak and not enough to overcome the electrostatic attraction from the macroion surface. On the other side, the electrostatic correlation between monovalent ions is weak too, and the entropic effect leads to a monotonic neutralization of the surface charge along the radial direction. Therefore, with the increase of the surface charge, both cation and anion
densities of monovalent salts remain monotonic curves.

For divalent ions, the situations are different. The strength of electric field is high near the surface. This produces the decrement of the dielectric permittivity as an output. The Born solvation energy is repulsive and has a quadratic relation with the ion valence. Although the surface-charge attraction is strong, a depletion zone can be observed for high surface charge densities $\sigma=-1.5$ and $-2.5~\textup{e/nm}^2$, demonstrating that the zone is difficult to access by mobile ions due to the repulsion from the Born solvation energy.
On the other hand, for these two surface densities, coion densities have a sharp increase up to a maximum at the one-ion-diameter distance, then decay to the bulk value monotonically. Observed from the distributions of electric potentials, the curves reverse the sign at the nearby distance, predicting a positive potential at the far field. This is the result of the correlation.
Strange
charge inversion phenomenon for multivalent counterions has been confirmed theoretically \cite{Shklovskii:PRE:1999,QGM+:CPC:2003,DL:JCP:2006} and experimentally by electrophoresis mobility \cite{MQG:JPCM:03,QGM:CSA:05,SHW:JCIS:11}. This clearly demonstrates again that the SCF model describes the electrostatic correlation. Ignoring the many-body correlation, the classical PB theory does not predict the charge inversion and as shown in Fig. \ref{fig:11symphiden}(f) that the electric potentials are always negative with the increase of $\sigma$.

\subsubsection{Effects of the field-dependent permittivity}

\begin{figure}[t!]
\includegraphics[scale=.36]{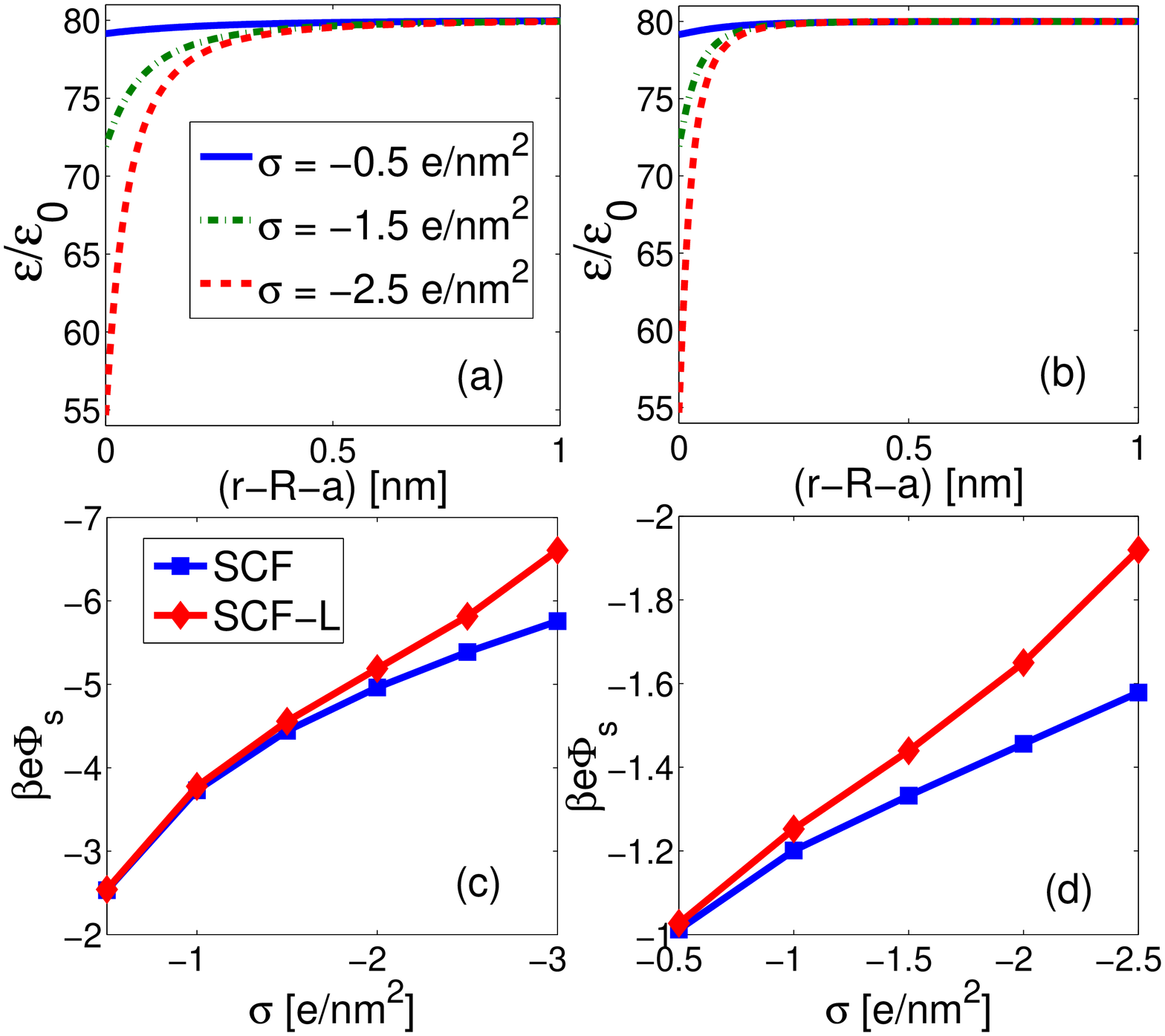}
\caption{The relative dielectric profiles as functions of radial distance to the surface in (a) 1:1  and (b) 2:1 symmetric electrolytes at room temperature $\ell_B = 0.714~\textup{nm}$ with three different surface charge densities. Comparison of surface potentials with SCF model (homogeneous water permittivity) and SCF-L model in: (c) 1:1 symmetric electrolyte and (d) 2:1 symmetric electrolyte.
}
\label{fig:21symdenvsPB}
\end{figure}

In Fig. \ref{fig:21symdenvsPB}(a)(b), the relative dielectric functions for 1:1 and 2:1 salts are illustrated in the cases of $\sigma=-0.5,~-1.5$ and $-2.5~\textup{e/nm}^2$. We shall see that the relative dielectric function is rapidly increased from $\sim 55$ to $>75$ for a distance less than one ion radius. In order to probe the effect of the dielectric variation in this small region, we present comparisons for the SCF calculations
between using this field-dependent dielectric function
and using uniform water dielectric constant $\varepsilon_W=80\varepsilon_0$. We call both results as the SCF-L and SCF, respectively.

Since the surface charge density plays an important role to provide different electric field and then affects the solvent structural details within the double layer \cite{FO:PCCP:11}, we calculate the surface potential as a function of the surface charge density and present the results in Fig \ref{fig:21symdenvsPB}(c)(d). The difference of surface potentials between two models becomes larger with the increase of $\sigma$, and significant differences appear in both 1:1 and 2:1 salts for large surface charges. For the 2:1 salt, the relative difference is already very high when $-\sigma$ is slightly bigger than $0.5~\textup{e}/\textup{nm}^2$ ($5\%$ for $\sigma=-1~\textup{e}/\textup{nm}^2$ and $14\%$ for $\sigma=-2~\textup{e}/\textup{nm}^2$). These results shall lead to the conclusion that the effect of field-dependent dielectric permittivity (i.e., the orientation of interfacial water molecules) are important for double-layer structure.

\subsubsection{Effect of the ionic size}

At last, let us see how the ionic size asymmetry affects the double-layer structure. We remain the anion radius of $a_-=0.225~\textup{nm}$, and consider two different cation (counterion) radii $a_+=0.225~\textup{nm}$ and $a_+=0.18~\textup{nm}$. The effect of ionic size is minor at the weak coupling regime. So we consider valence-symmetric electrolytes for the strong-coupling regime by setting the Bjerrum length $\ell_B=2.856~\textup{nm}$, a four times larger than the water Bjerrum length at room temperature. We can think this large Bjerrum length is because of a low temperature or a low bulk dielectric constant or multivalent ion species.

\begin{figure}[t!]
\includegraphics[scale=.32]{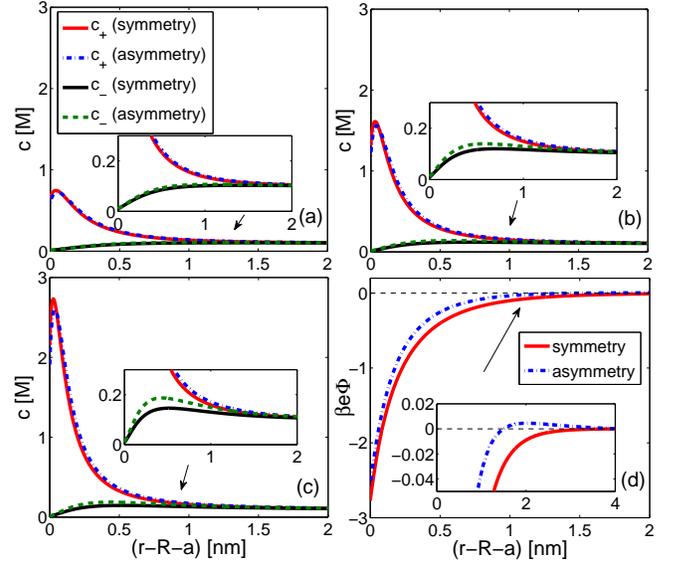}
\caption{Ion concentration and potential profiles for electrolytes of 1:1 salts with $\ell_B=2.856~\textup{nm}$.  Symmetric size ($a_+=a_-=0.225~\textup{nm}$) and asymmetric sizes ($a_+=0.18~\textup{nm}, a_-=0.225~\textup{nm}$). Ion concentrations for: (a) $\sigma=-0.2~\textup{e}/\textup{nm}^2$; (b) $\sigma=-0.3~\textup{e}/\textup{nm}^2$; and (c) $\sigma=-0.4~\textup{e}/\textup{nm}^2$.  (d) The electric potentials for $\sigma=-0.4~\textup{e}/\textup{nm}^2$.
} \label{fig:11Asymphi}
\end{figure}

Fig. \ref{fig:11Asymphi} plots the curves of both symmetric and asymmetric cases with three surface charge densities $\sigma=-0.2, -0.3$ and $-0.4~\textup{e}/\textup{nm}^2$. We shall note that for a large $\ell_B$, the electrostatic correlations between ions are strong and thus the nonmonotonic concentration should be predicted when $\sigma$ is not very low. On the other hand, the smaller counterion radius greatly increases the correlation self energy which is negative and thus is favorable for the ion congregation in the double layer. Both counterion and coion nearby densities of the asymmetric case (smaller counterions) are higher than the symmetric case, and this effect is enlarged with the increase of the surface charge density, as is shown in the insets. Fig. \ref{fig:11Asymphi}(d) shows the electric potentials in the case of $\sigma=-0.4~\textup{e}/\textup{nm}^2$ for symmetric and asymmetric ionic sizes. The coupling parameter is $\Xi=20.50$. We observed the charge inversion for the asymmetric case other from the symmetric case, illustrating the ionic sizes are very much relevant to ion correlations.

\section{Conclusions}

We have introduced a modified version of the self-consistent field model for electrostatic interaction, and developed a hybrid numerical method for the numerical solution of the resulted system of PDEs. In particular,
we proposed a new and efficient way to deal with the solution of self Green's function for a test ion as a charged hard sphere in a dielectrically inhomogeneous medium. As a result, the expensive direct numerical implementation for a multiscale (macroion and small ions) and high-dimensional Green's function problem (i.e., 6D problem for a 3D geometry) is avoided.

The numerical results are shown to capture many-body effects, such as the effect of inhomogeneous dielectric permittivity (the Born energy), size effect and ionic correlations. These effects are thought to be significant in many biological and physical phenomena near aqueous electrolyte interfaces. Strong ionic correlations lead to the charge inversion which has also been well reproduced by the SCF model. The numerical results are shown to quantitatively match the results from MC simulations. We shall notice the drawback of the approximation with Born series is obvious for irregular surfaces and remain the challenges of overcoming this issue by developing fast and accurate numerical methods as our future work. The effect of correlation to charge dynamics shall be an important direction to be addressed in the future, which will be studied in the framework of modified Poisson-Nernst-Planck equations.

\section*{Acknowledgement}
The authors acknowledge the support from the Natural Science Foundation of China (Grant Numbers: 11101276, and 91130012) and the Central Organization Department of China, and the HPC Center of SJTU. The authors thank Professor Wei Cai and Professor Xiangjun Xing for helpful suggestions.

%\bibliographystyle{elsart-num} %{elsart-num-sort} {plain} %
%\bibliography{C:/Users/user/Dropbox/XuGroup_Share/groupbib}  % for Manman desktop editing
%\bibliography{C:/Users/manmanma/Dropbox/XuGroup_Share/groupbib}  % for Manman Mac editing
%\bibliography{C:/Users/zhenli/Dropbox/XuGroup_Share/groupbib}    % for Zhenli desktop editing
%\bibliography{D:/Dropbox/XuGroup_Share/groupbib}  % for Zhenli laptop editing

\begin{thebibliography}{88}
\expandafter\ifx\csname natexlab\endcsname\relax\def\natexlab#1{#1}\fi
\expandafter\ifx\csname bibnamefont\endcsname\relax
  \def\bibnamefont#1{#1}\fi
\expandafter\ifx\csname bibfnamefont\endcsname\relax
  \def\bibfnamefont#1{#1}\fi
\expandafter\ifx\csname citenamefont\endcsname\relax
  \def\citenamefont#1{#1}\fi
\expandafter\ifx\csname url\endcsname\relax
  \def\url#1{\texttt{#1}}\fi
\expandafter\ifx\csname urlprefix\endcsname\relax\def\urlprefix{URL }\fi
\providecommand{\bibinfo}[2]{#2}
\providecommand{\eprint}[2][]{\url{#2}}

\bibitem[{\citenamefont{Levin}(2002)}]{Levin:RPP:2002}
\bibinfo{author}{\bibfnamefont{Y.}~\bibnamefont{Levin}}, \bibinfo{journal}{Rep.
  Prog. Phys.} \textbf{\bibinfo{volume}{65}}, \bibinfo{pages}{1577}
  (\bibinfo{year}{2002}).

\bibitem[{\citenamefont{Hunter}(1981)}]{Hunter::1981}
\bibinfo{author}{\bibfnamefont{R.~J.} \bibnamefont{Hunter}},
  \emph{\bibinfo{title}{Zeta Potential in Colloidal Sciences: Principles and
  Applications}} (\bibinfo{publisher}{Academic, Lundon}, \bibinfo{year}{1981}).

\bibitem[{\citenamefont{Joly et~al.}(2004)\citenamefont{Joly, Ybert, Trizac,
  and Bocquet}}]{JYTB:PRL:04}
\bibinfo{author}{\bibfnamefont{L.}~\bibnamefont{Joly}},
  \bibinfo{author}{\bibfnamefont{C.}~\bibnamefont{Ybert}},
  \bibinfo{author}{\bibfnamefont{E.}~\bibnamefont{Trizac}}, \bibnamefont{and}
  \bibinfo{author}{\bibfnamefont{L.}~\bibnamefont{Bocquet}},
  \bibinfo{journal}{Phys. Rev. Lett.} \textbf{\bibinfo{volume}{93}},
  \bibinfo{pages}{257805} (\bibinfo{year}{2004}).

\bibitem[{\citenamefont{French et~al.}(2010)\citenamefont{French, Parsegian,
  Podgornik, Rajter, Jagota, Luo, Asthagiri, Chaudhury, Chiang, Granick
  et~al.}}]{FPP+:RMP:2010}
\bibinfo{author}{\bibfnamefont{R.~H.} \bibnamefont{French}},
  \bibinfo{author}{\bibfnamefont{V.~A.} \bibnamefont{Parsegian}},
  \bibinfo{author}{\bibfnamefont{R.}~\bibnamefont{Podgornik}},
  \bibinfo{author}{\bibfnamefont{R.~F.} \bibnamefont{Rajter}},
  \bibinfo{author}{\bibfnamefont{A.}~\bibnamefont{Jagota}},
  \bibinfo{author}{\bibfnamefont{J.}~\bibnamefont{Luo}},
  \bibinfo{author}{\bibfnamefont{D.}~\bibnamefont{Asthagiri}},
  \bibinfo{author}{\bibfnamefont{M.~K.} \bibnamefont{Chaudhury}},
  \bibinfo{author}{\bibfnamefont{Y.-M.} \bibnamefont{Chiang}},
  \bibinfo{author}{\bibfnamefont{S.}~\bibnamefont{Granick}},
  \bibnamefont{et~al.}, \bibinfo{journal}{Rev. Mod. Phys.}
  \textbf{\bibinfo{volume}{82}}, \bibinfo{pages}{1887} (\bibinfo{year}{2010}).

\bibitem[{\citenamefont{Walker et~al.}(2011)\citenamefont{Walker, Kowalczyk,
  {de la Cruz}, and Grzybowski}}]{WK+:N:2011}
\bibinfo{author}{\bibfnamefont{D.~A.} \bibnamefont{Walker}},
  \bibinfo{author}{\bibfnamefont{B.}~\bibnamefont{Kowalczyk}},
  \bibinfo{author}{\bibfnamefont{M.~O.} \bibnamefont{{de la Cruz}}},
  \bibnamefont{and} \bibinfo{author}{\bibfnamefont{B.~A.}
  \bibnamefont{Grzybowski}}, \bibinfo{journal}{Nanoscale}
  \textbf{\bibinfo{volume}{3}}, \bibinfo{pages}{1316} (\bibinfo{year}{2011}).

\bibitem[{\citenamefont{Besteman et~al.}(2004)\citenamefont{Besteman,
  Zevenbergen, Heering, and Lemay}}]{BZH+:PRL:2004}
\bibinfo{author}{\bibfnamefont{K.}~\bibnamefont{Besteman}},
  \bibinfo{author}{\bibfnamefont{M.~A.~G.} \bibnamefont{Zevenbergen}},
  \bibinfo{author}{\bibfnamefont{H.~A.} \bibnamefont{Heering}},
  \bibnamefont{and} \bibinfo{author}{\bibfnamefont{S.~G.} \bibnamefont{Lemay}},
  \bibinfo{journal}{Phys. Rev. Lett.} \textbf{\bibinfo{volume}{93}},
  \bibinfo{pages}{170802} (\bibinfo{year}{2004}).

\bibitem[{\citenamefont{Pittler et~al.}(2006)\citenamefont{Pittler, Bu, Vaknin,
  Travesset, McGillivray, and L\"osche}}]{PBV+:PRL:2006}
\bibinfo{author}{\bibfnamefont{J.}~\bibnamefont{Pittler}},
  \bibinfo{author}{\bibfnamefont{W.}~\bibnamefont{Bu}},
  \bibinfo{author}{\bibfnamefont{D.}~\bibnamefont{Vaknin}},
  \bibinfo{author}{\bibfnamefont{A.}~\bibnamefont{Travesset}},
  \bibinfo{author}{\bibfnamefont{D.~J.} \bibnamefont{McGillivray}},
  \bibnamefont{and} \bibinfo{author}{\bibfnamefont{M.}~\bibnamefont{L\"osche}},
  \bibinfo{journal}{Phys. Rev. Lett.} \textbf{\bibinfo{volume}{97}},
  \bibinfo{pages}{046102} (\bibinfo{year}{2006}).

\bibitem[{\citenamefont{Lozada-Cassou et~al.}(1982)\citenamefont{Lozada-Cassou,
  Saavedra-Barrera, and Henderson}}]{LSH:JCP:82}
\bibinfo{author}{\bibfnamefont{M.}~\bibnamefont{Lozada-Cassou}},
  \bibinfo{author}{\bibfnamefont{R.}~\bibnamefont{Saavedra-Barrera}},
  \bibnamefont{and}
  \bibinfo{author}{\bibfnamefont{D.}~\bibnamefont{Henderson}},
  \bibinfo{journal}{J. Chem. Phys.} \textbf{\bibinfo{volume}{77}},
  \bibinfo{pages}{5150} (\bibinfo{year}{1982}).

\bibitem[{\citenamefont{Deserno et~al.}(2001)\citenamefont{Deserno,
  Jim{\'e}nez-{\'A}ngeles, Holm, and Lozada-Cassou}}]{DJH+:JPCB:2001}
\bibinfo{author}{\bibfnamefont{M.}~\bibnamefont{Deserno}},
  \bibinfo{author}{\bibfnamefont{F.}~\bibnamefont{Jim{\'e}nez-{\'A}ngeles}},
  \bibinfo{author}{\bibfnamefont{C.}~\bibnamefont{Holm}}, \bibnamefont{and}
  \bibinfo{author}{\bibfnamefont{M.}~\bibnamefont{Lozada-Cassou}},
  \bibinfo{journal}{J. Phys. Chem. B} \textbf{\bibinfo{volume}{105}},
  \bibinfo{pages}{10983} (\bibinfo{year}{2001}).

\bibitem[{\citenamefont{Grosberg et~al.}(2002)\citenamefont{Grosberg, Nguyen,
  and Shklovskii}}]{GNS:RMP:2002}
\bibinfo{author}{\bibfnamefont{A.~Y.} \bibnamefont{Grosberg}},
  \bibinfo{author}{\bibfnamefont{T.~T.} \bibnamefont{Nguyen}},
  \bibnamefont{and} \bibinfo{author}{\bibfnamefont{B.~I.}
  \bibnamefont{Shklovskii}}, \bibinfo{journal}{Rev. Mod. Phys.}
  \textbf{\bibinfo{volume}{74}}, \bibinfo{pages}{329} (\bibinfo{year}{2002}).

\bibitem[{\citenamefont{Larsen and Grier}(1997)}]{LG:N:1997}
\bibinfo{author}{\bibfnamefont{A.~E.} \bibnamefont{Larsen}} \bibnamefont{and}
  \bibinfo{author}{\bibfnamefont{D.~G.} \bibnamefont{Grier}},
  \bibinfo{journal}{Nature} \textbf{\bibinfo{volume}{385}},
  \bibinfo{pages}{230} (\bibinfo{year}{1997}).

\bibitem[{\citenamefont{Angelini et~al.}(2003)\citenamefont{Angelini, Liang,
  Wriggers, and Wong}}]{ALW+:PNASU:2003}
\bibinfo{author}{\bibfnamefont{T.~E.} \bibnamefont{Angelini}},
  \bibinfo{author}{\bibfnamefont{H.}~\bibnamefont{Liang}},
  \bibinfo{author}{\bibfnamefont{W.}~\bibnamefont{Wriggers}}, \bibnamefont{and}
  \bibinfo{author}{\bibfnamefont{G.~C.~L.} \bibnamefont{Wong}},
  \bibinfo{journal}{Proc. Nat. Acad. Sci. USA} \textbf{\bibinfo{volume}{100}},
  \bibinfo{pages}{8634} (\bibinfo{year}{2003}).

\bibitem[{\citenamefont{Eisenberg}(2011)}]{Eisenberg:ACPip:2011a}
\bibinfo{author}{\bibfnamefont{B.}~\bibnamefont{Eisenberg}},
  \bibinfo{journal}{Adv. Chem. Phys.} \textbf{\bibinfo{volume}{31}},
  \bibinfo{pages}{117} (\bibinfo{year}{2011}).

\bibitem[{\citenamefont{Fedorov and Kornyshev}(2014)}]{FK:CR:14}
\bibinfo{author}{\bibfnamefont{M.~V.} \bibnamefont{Fedorov}} \bibnamefont{and}
  \bibinfo{author}{\bibfnamefont{A.~A.} \bibnamefont{Kornyshev}},
  \bibinfo{journal}{Chem. Rev.} \textbf{\bibinfo{volume}{114}},
  \bibinfo{pages}{2978} (\bibinfo{year}{2014}).

\bibitem[{\citenamefont{Boroudjerdi et~al.}(2005)\citenamefont{Boroudjerdi,
  Kim, Naji, Netz, Schlagberger, and Serr}}]{BKN+:PR:2005}
\bibinfo{author}{\bibfnamefont{H.}~\bibnamefont{Boroudjerdi}},
  \bibinfo{author}{\bibfnamefont{Y.-W.} \bibnamefont{Kim}},
  \bibinfo{author}{\bibfnamefont{A.}~\bibnamefont{Naji}},
  \bibinfo{author}{\bibfnamefont{R.~R.} \bibnamefont{Netz}},
  \bibinfo{author}{\bibfnamefont{X.}~\bibnamefont{Schlagberger}},
  \bibnamefont{and} \bibinfo{author}{\bibfnamefont{A.}~\bibnamefont{Serr}},
  \bibinfo{journal}{Phys. Rep.} \textbf{\bibinfo{volume}{416}},
  \bibinfo{pages}{129} (\bibinfo{year}{2005}).

\bibitem[{\citenamefont{Borukhov et~al.}(1997)\citenamefont{Borukhov, Andelman,
  and Orland}}]{BAO:PRL:1997}
\bibinfo{author}{\bibfnamefont{I.}~\bibnamefont{Borukhov}},
  \bibinfo{author}{\bibfnamefont{D.}~\bibnamefont{Andelman}}, \bibnamefont{and}
  \bibinfo{author}{\bibfnamefont{H.}~\bibnamefont{Orland}},
  \bibinfo{journal}{Phys. Rev. Lett.} \textbf{\bibinfo{volume}{79}},
  \bibinfo{pages}{435} (\bibinfo{year}{1997}).

\bibitem[{\citenamefont{Kilic et~al.}(2007)\citenamefont{Kilic, Bazant, and
  Ajdari}}]{KBA:PRE:2007}
\bibinfo{author}{\bibfnamefont{M.~S.} \bibnamefont{Kilic}},
  \bibinfo{author}{\bibfnamefont{M.~Z.} \bibnamefont{Bazant}},
  \bibnamefont{and} \bibinfo{author}{\bibfnamefont{A.}~\bibnamefont{Ajdari}},
  \bibinfo{journal}{Phys. Rev. E} \textbf{\bibinfo{volume}{75}},
  \bibinfo{pages}{021503} (\bibinfo{year}{2007}).

\bibitem[{\citenamefont{Corry et~al.}(2003)\citenamefont{Corry, Kuyucak, and
  Chung}}]{CKC:BJ:03}
\bibinfo{author}{\bibfnamefont{B.}~\bibnamefont{Corry}},
  \bibinfo{author}{\bibfnamefont{S.}~\bibnamefont{Kuyucak}}, \bibnamefont{and}
  \bibinfo{author}{\bibfnamefont{S.-H.} \bibnamefont{Chung}},
  \bibinfo{journal}{Biophys. J.} \textbf{\bibinfo{volume}{84}},
  \bibinfo{pages}{3594} (\bibinfo{year}{2003}).

\bibitem[{\citenamefont{Graf et~al.}(2004)\citenamefont{Graf, Kurnikova,
  Coalson, and Nitzan}}]{GKCN:JPC:04}
\bibinfo{author}{\bibfnamefont{P.}~\bibnamefont{Graf}},
  \bibinfo{author}{\bibfnamefont{M.~G.} \bibnamefont{Kurnikova}},
  \bibinfo{author}{\bibfnamefont{R.~D.} \bibnamefont{Coalson}},
  \bibnamefont{and} \bibinfo{author}{\bibfnamefont{A.}~\bibnamefont{Nitzan}},
  \bibinfo{journal}{J. Phys. Chem. B} \textbf{\bibinfo{volume}{108}},
  \bibinfo{pages}{2006} (\bibinfo{year}{2004}).

\bibitem[{\citenamefont{Eisenberg et~al.}(2010)\citenamefont{Eisenberg, Hyon,
  and Liu}}]{EHL:JCP:2010}
\bibinfo{author}{\bibfnamefont{B.}~\bibnamefont{Eisenberg}},
  \bibinfo{author}{\bibfnamefont{Y.}~\bibnamefont{Hyon}}, \bibnamefont{and}
  \bibinfo{author}{\bibfnamefont{C.}~\bibnamefont{Liu}}, \bibinfo{journal}{J.
  Chem. Phys.} \textbf{\bibinfo{volume}{133}}, \bibinfo{pages}{104104}
  (\bibinfo{year}{2010}).

\bibitem[{\citenamefont{Hansen and McDonald}(2006)}]{HM::2006}
\bibinfo{author}{\bibfnamefont{J.~P.} \bibnamefont{Hansen}} \bibnamefont{and}
  \bibinfo{author}{\bibfnamefont{I.~R.} \bibnamefont{McDonald}},
  \emph{\bibinfo{title}{Theory of simple liquids}}
  (\bibinfo{publisher}{Academic Press}, \bibinfo{address}{Amsterdam},
  \bibinfo{year}{2006}).

\bibitem[{\citenamefont{Bazant et~al.}(2011)\citenamefont{Bazant, Storey, and
  Kornyshev}}]{BSK:PRL:2011}
\bibinfo{author}{\bibfnamefont{M.~Z.} \bibnamefont{Bazant}},
  \bibinfo{author}{\bibfnamefont{B.~D.} \bibnamefont{Storey}},
  \bibnamefont{and} \bibinfo{author}{\bibfnamefont{A.~A.}
  \bibnamefont{Kornyshev}}, \bibinfo{journal}{Phys. Rev. Lett.}
  \textbf{\bibinfo{volume}{106}}, \bibinfo{pages}{046102}
  (\bibinfo{year}{2011}).

\bibitem[{\citenamefont{Liu and Eisenberg}(2013)}]{LE:JPCB:13}
\bibinfo{author}{\bibfnamefont{J.-L.} \bibnamefont{Liu}} \bibnamefont{and}
  \bibinfo{author}{\bibfnamefont{B.}~\bibnamefont{Eisenberg}},
  \bibinfo{journal}{J. Phys. Chem. B} \textbf{\bibinfo{volume}{117}},
  \bibinfo{pages}{12051} (\bibinfo{year}{2013}).

\bibitem[{\citenamefont{Linse}(2005)}]{Linse:APS:2005}
\bibinfo{author}{\bibfnamefont{P.}~\bibnamefont{Linse}}, \bibinfo{journal}{Adv.
  Polym. Sci.} \textbf{\bibinfo{volume}{185}}, \bibinfo{pages}{111}
  (\bibinfo{year}{2005}).

\bibitem[{\citenamefont{Messina}(2009)}]{Messina:JPCM:2009}
\bibinfo{author}{\bibfnamefont{R.}~\bibnamefont{Messina}}, \bibinfo{journal}{J.
  Phys. Condens. Matter} \textbf{\bibinfo{volume}{21}}, \bibinfo{pages}{113102}
  (\bibinfo{year}{2009}).

\bibitem[{\citenamefont{Hasted et~al.}(1948)\citenamefont{Hasted, Ritson, and
  Collie}}]{HRC:JCP:48}
\bibinfo{author}{\bibfnamefont{J.}~\bibnamefont{Hasted}},
  \bibinfo{author}{\bibfnamefont{D.}~\bibnamefont{Ritson}}, \bibnamefont{and}
  \bibinfo{author}{\bibfnamefont{C.}~\bibnamefont{Collie}},
  \bibinfo{journal}{J. Chem. Phys.} \textbf{\bibinfo{volume}{16}},
  \bibinfo{pages}{1} (\bibinfo{year}{1948}).

\bibitem[{\citenamefont{Booth}(1951)}]{B:JCP:1951}
\bibinfo{author}{\bibfnamefont{F.}~\bibnamefont{Booth}}, \bibinfo{journal}{J.
  Chem. Phys.} \textbf{\bibinfo{volume}{19}}, \bibinfo{pages}{391}
  (\bibinfo{year}{1951}).

\bibitem[{\citenamefont{Booth}(1955)}]{B:JCP:1955}
\bibinfo{author}{\bibfnamefont{F.}~\bibnamefont{Booth}}, \bibinfo{journal}{J.
  Chem. Phys.} \textbf{\bibinfo{volume}{23}}, \bibinfo{pages}{453}
  (\bibinfo{year}{1955}).

\bibitem[{\citenamefont{Yeh and Berkowitz}(1999)}]{YB:JCP:99}
\bibinfo{author}{\bibfnamefont{I.-C.} \bibnamefont{Yeh}} \bibnamefont{and}
  \bibinfo{author}{\bibfnamefont{M.~L.} \bibnamefont{Berkowitz}},
  \bibinfo{journal}{J. Chem. Phys.} \textbf{\bibinfo{volume}{110}},
  \bibinfo{pages}{7935} (\bibinfo{year}{1999}).

\bibitem[{\citenamefont{Fulton}(2009)}]{Fulton:JCP:09}
\bibinfo{author}{\bibfnamefont{R.~L.} \bibnamefont{Fulton}},
  \bibinfo{journal}{J. Chem. Phys.} \textbf{\bibinfo{volume}{130}},
  \bibinfo{pages}{204503} (\bibinfo{year}{2009}).

\bibitem[{\citenamefont{Gur et~al.}(1978)\citenamefont{Gur, Ravina, and
  Babchin}}]{GRB:JCIS:78}
\bibinfo{author}{\bibfnamefont{Y.}~\bibnamefont{Gur}},
  \bibinfo{author}{\bibfnamefont{I.}~\bibnamefont{Ravina}}, \bibnamefont{and}
  \bibinfo{author}{\bibfnamefont{A.~J.} \bibnamefont{Babchin}},
  \bibinfo{journal}{J. Colloid Interf. Sci.} \textbf{\bibinfo{volume}{64}},
  \bibinfo{pages}{333} (\bibinfo{year}{1978}).

\bibitem[{\citenamefont{Paunov et~al.}(1996)\citenamefont{Paunov, Dimova,
  Kralchevsky, Broze, and Mehreteab}}]{PDKBM:JCIS:96}
\bibinfo{author}{\bibfnamefont{V.}~\bibnamefont{Paunov}},
  \bibinfo{author}{\bibfnamefont{R.}~\bibnamefont{Dimova}},
  \bibinfo{author}{\bibfnamefont{P.}~\bibnamefont{Kralchevsky}},
  \bibinfo{author}{\bibfnamefont{G.}~\bibnamefont{Broze}}, \bibnamefont{and}
  \bibinfo{author}{\bibfnamefont{A.}~\bibnamefont{Mehreteab}},
  \bibinfo{journal}{J. Colloid Interf. Sci.} \textbf{\bibinfo{volume}{182}},
  \bibinfo{pages}{239} (\bibinfo{year}{1996}).

\bibitem[{\citenamefont{Ben-Yaakov et~al.}(2011)\citenamefont{Ben-Yaakov,
  Andelman, and Podgornik}}]{BAP:JCP:11}
\bibinfo{author}{\bibfnamefont{D.}~\bibnamefont{Ben-Yaakov}},
  \bibinfo{author}{\bibfnamefont{D.}~\bibnamefont{Andelman}}, \bibnamefont{and}
  \bibinfo{author}{\bibfnamefont{R.}~\bibnamefont{Podgornik}},
  \bibinfo{journal}{J. Chem. Phys.} \textbf{\bibinfo{volume}{134}},
  \bibinfo{pages}{074705} (\bibinfo{year}{2011}).

\bibitem[{\citenamefont{Wang et~al.}(2011)\citenamefont{Wang, Varghese, and
  Pilon}}]{WVP:EA:11}
\bibinfo{author}{\bibfnamefont{H.}~\bibnamefont{Wang}},
  \bibinfo{author}{\bibfnamefont{J.}~\bibnamefont{Varghese}}, \bibnamefont{and}
  \bibinfo{author}{\bibfnamefont{L.}~\bibnamefont{Pilon}},
  \bibinfo{journal}{Electrochimica Acta} \textbf{\bibinfo{volume}{56}},
  \bibinfo{pages}{6189} (\bibinfo{year}{2011}).

\bibitem[{\citenamefont{Frydel and Oettel}(2011)}]{FO:PCCP:11}
\bibinfo{author}{\bibfnamefont{D.}~\bibnamefont{Frydel}} \bibnamefont{and}
  \bibinfo{author}{\bibfnamefont{M.}~\bibnamefont{Oettel}},
  \bibinfo{journal}{Phys. Chem. Chem. Phys.} \textbf{\bibinfo{volume}{13}},
  \bibinfo{pages}{4109} (\bibinfo{year}{2011}).

\bibitem[{\citenamefont{Bonthuis et~al.}(2011)\citenamefont{Bonthuis, Gekle,
  and Netz}}]{BGN:PRL:2011}
\bibinfo{author}{\bibfnamefont{D.~J.} \bibnamefont{Bonthuis}},
  \bibinfo{author}{\bibfnamefont{S.}~\bibnamefont{Gekle}}, \bibnamefont{and}
  \bibinfo{author}{\bibfnamefont{R.~R.} \bibnamefont{Netz}},
  \bibinfo{journal}{Phys. Rev. Lett.} \textbf{\bibinfo{volume}{107}},
  \bibinfo{pages}{166102} (\bibinfo{year}{2011}).

\bibitem[{\citenamefont{Bonthuis et~al.}(2012)\citenamefont{Bonthuis, Gekle,
  and Netz}}]{BGN:Langmuir:12}
\bibinfo{author}{\bibfnamefont{D.~J.} \bibnamefont{Bonthuis}},
  \bibinfo{author}{\bibfnamefont{S.}~\bibnamefont{Gekle}}, \bibnamefont{and}
  \bibinfo{author}{\bibfnamefont{R.~R.} \bibnamefont{Netz}},
  \bibinfo{journal}{Langmuir} \textbf{\bibinfo{volume}{28}},
  \bibinfo{pages}{7679} (\bibinfo{year}{2012}).

\bibitem[{\citenamefont{Bonthuis and Netz}(2013)}]{BN:JPC:13}
\bibinfo{author}{\bibfnamefont{D.~J.} \bibnamefont{Bonthuis}} \bibnamefont{and}
  \bibinfo{author}{\bibfnamefont{R.~R.} \bibnamefont{Netz}},
  \bibinfo{journal}{J. Phys. Chem. B} \textbf{\bibinfo{volume}{117}},
  \bibinfo{pages}{11397} (\bibinfo{year}{2013}).

\bibitem[{\citenamefont{Onsager and Samaras}(1934)}]{OS:JCP:1934}
\bibinfo{author}{\bibfnamefont{L.}~\bibnamefont{Onsager}} \bibnamefont{and}
  \bibinfo{author}{\bibfnamefont{N.~N.~T.} \bibnamefont{Samaras}},
  \bibinfo{journal}{J. Chem. Phys.} \textbf{\bibinfo{volume}{2}},
  \bibinfo{pages}{528} (\bibinfo{year}{1934}).

\bibitem[{\citenamefont{Podgornik}(1989)}]{podgornik1989jcp}
\bibinfo{author}{\bibfnamefont{R.}~\bibnamefont{Podgornik}},
  \bibinfo{journal}{J. Chem. Phys.} \textbf{\bibinfo{volume}{91}},
  \bibinfo{pages}{5840} (\bibinfo{year}{1989}).

\bibitem[{\citenamefont{Netz and Orland}(2000)}]{NO:EPJE:2000}
\bibinfo{author}{\bibfnamefont{R.~R.} \bibnamefont{Netz}} \bibnamefont{and}
  \bibinfo{author}{\bibfnamefont{H.}~\bibnamefont{Orland}},
  \bibinfo{journal}{Eur. Phys. J. E} \textbf{\bibinfo{volume}{1}},
  \bibinfo{pages}{203} (\bibinfo{year}{2000}).

\bibitem[{\citenamefont{Netz and Orland}(2003)}]{NO:EPJE:2003}
\bibinfo{author}{\bibfnamefont{R.~R.} \bibnamefont{Netz}} \bibnamefont{and}
  \bibinfo{author}{\bibfnamefont{H.}~\bibnamefont{Orland}},
  \bibinfo{journal}{Eur. Phys. J. E} \textbf{\bibinfo{volume}{11}},
  \bibinfo{pages}{301} (\bibinfo{year}{2003}).

\bibitem[{\citenamefont{Wang}(2010)}]{Wang:PRE:2010}
\bibinfo{author}{\bibfnamefont{Z.~G.} \bibnamefont{Wang}},
  \bibinfo{journal}{Phys. Rev. E} \textbf{\bibinfo{volume}{81}},
  \bibinfo{pages}{021501} (\bibinfo{year}{2010}).

\bibitem[{\citenamefont{Wang and Wang}(2013)}]{WangRui:JCP:13}
\bibinfo{author}{\bibfnamefont{R.}~\bibnamefont{Wang}} \bibnamefont{and}
  \bibinfo{author}{\bibfnamefont{Z.-G.} \bibnamefont{Wang}},
  \bibinfo{journal}{J. Chem. Phys.} \textbf{\bibinfo{volume}{139}},
  \bibinfo{pages}{124702} (\bibinfo{year}{2013}).

\bibitem[{\citenamefont{Lu and Xing}(2014)}]{LX:PRE:2014}
\bibinfo{author}{\bibfnamefont{B.-S.} \bibnamefont{Lu}} \bibnamefont{and}
  \bibinfo{author}{\bibfnamefont{X.}~\bibnamefont{Xing}},
  \bibinfo{journal}{Phys. Rev. E} \textbf{\bibinfo{volume}{89}},
  \bibinfo{pages}{032305} (\bibinfo{year}{2014}).

\bibitem[{\citenamefont{Buyukdagli et~al.}(2012)\citenamefont{Buyukdagli,
  Achim, and Ala-Nissila}}]{BAA:JCP:2012}
\bibinfo{author}{\bibfnamefont{S.}~\bibnamefont{Buyukdagli}},
  \bibinfo{author}{\bibfnamefont{C.~V.} \bibnamefont{Achim}}, \bibnamefont{and}
  \bibinfo{author}{\bibfnamefont{T.}~\bibnamefont{Ala-Nissila}},
  \bibinfo{journal}{J. Chem. Phys.} \textbf{\bibinfo{volume}{137}},
  \bibinfo{pages}{104902} (\bibinfo{year}{2012}).

\bibitem[{\citenamefont{Lau et~al.}(2002)\citenamefont{Lau, Lukatsky, Pincus,
  and Safran}}]{LLP+:PRE:2002}
\bibinfo{author}{\bibfnamefont{A.~W.~C.} \bibnamefont{Lau}},
  \bibinfo{author}{\bibfnamefont{D.~B.} \bibnamefont{Lukatsky}},
  \bibinfo{author}{\bibfnamefont{P.}~\bibnamefont{Pincus}}, \bibnamefont{and}
  \bibinfo{author}{\bibfnamefont{S.~A.} \bibnamefont{Safran}},
  \bibinfo{journal}{Phys. Rev. E} \textbf{\bibinfo{volume}{65}},
  \bibinfo{pages}{051502} (\bibinfo{year}{2002}).

\bibitem[{\citenamefont{Born}(1920)}]{Born:ZP:1920}
\bibinfo{author}{\bibfnamefont{M.}~\bibnamefont{Born}}, \bibinfo{journal}{Z.
  Phys.} \textbf{\bibinfo{volume}{1}}, \bibinfo{pages}{45}
  (\bibinfo{year}{1920}).

\bibitem[{\citenamefont{Xu and Maggs}(2014)}]{XuMaggs:JCP:14}
\bibinfo{author}{\bibfnamefont{Z.}~\bibnamefont{Xu}} \bibnamefont{and}
  \bibinfo{author}{\bibfnamefont{A.}~\bibnamefont{Maggs}}, \bibinfo{journal}{J.
  Comput. Phys.} \textbf{\bibinfo{volume}{275}}, \bibinfo{pages}{310}
  (\bibinfo{year}{2014}).

\bibitem[{\citenamefont{Luo et~al.}(2006)\citenamefont{Luo, Malkova, Yoon,
  Schultz, Lin, Meron, Benjamin, Vanýsek, and Schlossman}}]{Luo06}
\bibinfo{author}{\bibfnamefont{G.}~\bibnamefont{Luo}},
  \bibinfo{author}{\bibfnamefont{S.}~\bibnamefont{Malkova}},
  \bibinfo{author}{\bibfnamefont{J.}~\bibnamefont{Yoon}},
  \bibinfo{author}{\bibfnamefont{D.~G.} \bibnamefont{Schultz}},
  \bibinfo{author}{\bibfnamefont{B.}~\bibnamefont{Lin}},
  \bibinfo{author}{\bibfnamefont{M.}~\bibnamefont{Meron}},
  \bibinfo{author}{\bibfnamefont{I.}~\bibnamefont{Benjamin}},
  \bibinfo{author}{\bibfnamefont{P.}~\bibnamefont{Vanýsek}}, \bibnamefont{and}
  \bibinfo{author}{\bibfnamefont{M.~L.} \bibnamefont{Schlossman}},
  \bibinfo{journal}{Science} \textbf{\bibinfo{volume}{311}},
  \bibinfo{pages}{216} (\bibinfo{year}{2006}).

\bibitem[{\citenamefont{Gillespie}(2014)}]{Gillespie:MN:14}
\bibinfo{author}{\bibfnamefont{D.}~\bibnamefont{Gillespie}},
  \bibinfo{journal}{Microfluidics and Nanofluidics} pp. \bibinfo{pages}{1--22}
  (\bibinfo{year}{2014}).

\bibitem[{\citenamefont{Li}(2009)}]{Li:N:2009}
\bibinfo{author}{\bibfnamefont{B.}~\bibnamefont{Li}},
  \bibinfo{journal}{Nonlinearity} \textbf{\bibinfo{volume}{22}},
  \bibinfo{pages}{811} (\bibinfo{year}{2009}).

\bibitem[{\citenamefont{Li et~al.}(2013)\citenamefont{Li, Liu, Xu, and
  Zhou}}]{LLXS:NonL:2013}
\bibinfo{author}{\bibfnamefont{B.}~\bibnamefont{Li}},
  \bibinfo{author}{\bibfnamefont{P.}~\bibnamefont{Liu}},
  \bibinfo{author}{\bibfnamefont{Z.}~\bibnamefont{Xu}}, \bibnamefont{and}
  \bibinfo{author}{\bibfnamefont{S.}~\bibnamefont{Zhou}},
  \bibinfo{journal}{Nonlinearity} \textbf{\bibinfo{volume}{26}},
  \bibinfo{pages}{2899} (\bibinfo{year}{2013}).

\bibitem[{\citenamefont{Frydel and Levin}(2012)}]{FL:JCP:12}
\bibinfo{author}{\bibfnamefont{D.}~\bibnamefont{Frydel}} \bibnamefont{and}
  \bibinfo{author}{\bibfnamefont{Y.}~\bibnamefont{Levin}}, \bibinfo{journal}{J.
  Chem. Phys.} \textbf{\bibinfo{volume}{137}}, \bibinfo{pages}{164703}
  (\bibinfo{year}{2012}).

\bibitem[{\citenamefont{Rosenfeld}(1989)}]{rosenfeld:PRL:1989}
\bibinfo{author}{\bibfnamefont{Y.}~\bibnamefont{Rosenfeld}},
  \bibinfo{journal}{Phys. Rev. Lett.} \textbf{\bibinfo{volume}{63}},
  \bibinfo{pages}{980} (\bibinfo{year}{1989}).

\bibitem[{\citenamefont{Xu et~al.}(2014)\citenamefont{Xu, Ma, and
  Liu}}]{XML:PRE:2014}
\bibinfo{author}{\bibfnamefont{Z.}~\bibnamefont{Xu}},
  \bibinfo{author}{\bibfnamefont{M.}~\bibnamefont{Ma}}, \bibnamefont{and}
  \bibinfo{author}{\bibfnamefont{P.}~\bibnamefont{Liu}},
  \bibinfo{journal}{Phys. Rev. E} \textbf{\bibinfo{volume}{90}},
  \bibinfo{pages}{013307} (\bibinfo{year}{2014}).

\bibitem[{\citenamefont{Levin et~al.}(2009)\citenamefont{Levin, Dos~Santos, and
  Diehl}}]{Levin:PRL1:09}
\bibinfo{author}{\bibfnamefont{Y.}~\bibnamefont{Levin}},
  \bibinfo{author}{\bibfnamefont{A.~P.} \bibnamefont{Dos~Santos}},
  \bibnamefont{and} \bibinfo{author}{\bibfnamefont{A.}~\bibnamefont{Diehl}},
  \bibinfo{journal}{Phys. Rev. Lett.} \textbf{\bibinfo{volume}{103}},
  \bibinfo{pages}{257802} (\bibinfo{year}{2009}).

\bibitem[{\citenamefont{Levin}(2009)}]{Levin:PRL2:09}
\bibinfo{author}{\bibfnamefont{Y.}~\bibnamefont{Levin}},
  \bibinfo{journal}{Phys. Rev. Lett.} \textbf{\bibinfo{volume}{102}},
  \bibinfo{pages}{147803} (\bibinfo{year}{2009}).

\bibitem[{\citenamefont{Wang and Wang}(2014)}]{WW:PRL:14}
\bibinfo{author}{\bibfnamefont{R.}~\bibnamefont{Wang}} \bibnamefont{and}
  \bibinfo{author}{\bibfnamefont{Z.-G.} \bibnamefont{Wang}},
  \bibinfo{journal}{Phys. Rev. Lett.} \textbf{\bibinfo{volume}{112}},
  \bibinfo{pages}{136101} (\bibinfo{year}{2014}).

\bibitem[{\citenamefont{Abrashkin et~al.}(2007)\citenamefont{Abrashkin,
  Andelman, and Orland}}]{AAO:PRL:2007}
\bibinfo{author}{\bibfnamefont{A.}~\bibnamefont{Abrashkin}},
  \bibinfo{author}{\bibfnamefont{D.}~\bibnamefont{Andelman}}, \bibnamefont{and}
  \bibinfo{author}{\bibfnamefont{H.}~\bibnamefont{Orland}},
  \bibinfo{journal}{Phys. Rev. Lett.} \textbf{\bibinfo{volume}{99}},
  \bibinfo{pages}{077801} (\bibinfo{year}{2007}).

\bibitem[{\citenamefont{Frydel}(2014)}]{Frydel:review:2014}
\bibinfo{author}{\bibfnamefont{D.}~\bibnamefont{Frydel}},
  \bibinfo{journal}{Personal Communication}  (\bibinfo{year}{2014}).

\bibitem[{\citenamefont{Bockris and Reddy}(1998)}]{BR:NY:98}
\bibinfo{author}{\bibfnamefont{J.}~\bibnamefont{Bockris}} \bibnamefont{and}
  \bibinfo{author}{\bibfnamefont{A.}~\bibnamefont{Reddy}},
  \emph{\bibinfo{title}{Modern {Electrochemistry}}} (\bibinfo{publisher}{Plenum
  Press, New York}, \bibinfo{year}{1998}).

\bibitem[{\citenamefont{Kalcher and Dzubiella}(2009)}]{KD:JCP:09}
\bibinfo{author}{\bibfnamefont{I.}~\bibnamefont{Kalcher}} \bibnamefont{and}
  \bibinfo{author}{\bibfnamefont{J.}~\bibnamefont{Dzubiella}},
  \bibinfo{journal}{J. Chem. Phys.} \textbf{\bibinfo{volume}{130}},
  \bibinfo{pages}{134507} (\bibinfo{year}{2009}).

\bibitem[{\citenamefont{Li et~al.}(2014)\citenamefont{Li, Wen, and
  Zhou}}]{LWZ:CMS:14}
\bibinfo{author}{\bibfnamefont{B.}~\bibnamefont{Li}},
  \bibinfo{author}{\bibfnamefont{J.}~\bibnamefont{Wen}}, \bibnamefont{and}
  \bibinfo{author}{\bibfnamefont{S.}~\bibnamefont{Zhou}},
  \bibinfo{journal}{Personal Communication}  (\bibinfo{year}{2014}).

\bibitem[{\citenamefont{Renou et~al.}(2014)\citenamefont{Renou, Ding, Zhu,
  Szymczyk, Malfreyt, and Ghoufi}}]{Renou:JPCB:14}
\bibinfo{author}{\bibfnamefont{R.}~\bibnamefont{Renou}},
  \bibinfo{author}{\bibfnamefont{M.}~\bibnamefont{Ding}},
  \bibinfo{author}{\bibfnamefont{H.}~\bibnamefont{Zhu}},
  \bibinfo{author}{\bibfnamefont{A.}~\bibnamefont{Szymczyk}},
  \bibinfo{author}{\bibfnamefont{P.}~\bibnamefont{Malfreyt}}, \bibnamefont{and}
  \bibinfo{author}{\bibfnamefont{A.}~\bibnamefont{Ghoufi}},
  \bibinfo{journal}{J. Phys. Chem. B} \textbf{\bibinfo{volume}{118}},
  \bibinfo{pages}{3931} (\bibinfo{year}{2014}).

\bibitem[{\citenamefont{Frydel}(2011)}]{Frydel:JCP:11}
\bibinfo{author}{\bibfnamefont{D.}~\bibnamefont{Frydel}}, \bibinfo{journal}{J.
  Chem. Phys.} \textbf{\bibinfo{volume}{134}}, \bibinfo{pages}{234704}
  (\bibinfo{year}{2011}).

\bibitem[{\citenamefont{Lin et~al.}(2011{\natexlab{a}})\citenamefont{Lin, Yang,
  Lu, Ying, and E}}]{LYL+:SJoSC:2011}
\bibinfo{author}{\bibfnamefont{L.}~\bibnamefont{Lin}},
  \bibinfo{author}{\bibfnamefont{C.}~\bibnamefont{Yang}},
  \bibinfo{author}{\bibfnamefont{J.}~\bibnamefont{Lu}},
  \bibinfo{author}{\bibfnamefont{L.}~\bibnamefont{Ying}}, \bibnamefont{and}
  \bibinfo{author}{\bibfnamefont{W.}~\bibnamefont{E}}, \bibinfo{journal}{SIAM
  J. Sci. Comput.} \textbf{\bibinfo{volume}{33}}, \bibinfo{pages}{1329}
  (\bibinfo{year}{2011}{\natexlab{a}}).

\bibitem[{\citenamefont{Lin et~al.}(2011{\natexlab{b}})\citenamefont{Lin, Yang,
  Meza, Lu, Ying, and E}}]{LYM+:ATMS:2011}
\bibinfo{author}{\bibfnamefont{L.}~\bibnamefont{Lin}},
  \bibinfo{author}{\bibfnamefont{C.}~\bibnamefont{Yang}},
  \bibinfo{author}{\bibfnamefont{J.~C.} \bibnamefont{Meza}},
  \bibinfo{author}{\bibfnamefont{J.}~\bibnamefont{Lu}},
  \bibinfo{author}{\bibfnamefont{L.}~\bibnamefont{Ying}}, \bibnamefont{and}
  \bibinfo{author}{\bibfnamefont{W.}~\bibnamefont{E}}, \bibinfo{journal}{ACM
  Trans. Math. Softw.} \textbf{\bibinfo{volume}{37}}, \bibinfo{pages}{40:1}
  (\bibinfo{year}{2011}{\natexlab{b}}).

\bibitem[{\citenamefont{Teschke et~al.}(2001)\citenamefont{Teschke, Ceotto, and
  De~Souza}}]{TCS:PRE:01}
\bibinfo{author}{\bibfnamefont{O.}~\bibnamefont{Teschke}},
  \bibinfo{author}{\bibfnamefont{G.}~\bibnamefont{Ceotto}}, \bibnamefont{and}
  \bibinfo{author}{\bibfnamefont{E.}~\bibnamefont{De~Souza}},
  \bibinfo{journal}{Phys. Rev. E} \textbf{\bibinfo{volume}{64}},
  \bibinfo{pages}{011605} (\bibinfo{year}{2001}).

\bibitem[{\citenamefont{Podgornik et~al.}(1987)\citenamefont{Podgornik, Cevc,
  and {\v{Z}}ek{\v{s}}}}]{PCZ:JCP:87}
\bibinfo{author}{\bibfnamefont{R.}~\bibnamefont{Podgornik}},
  \bibinfo{author}{\bibfnamefont{G.}~\bibnamefont{Cevc}}, \bibnamefont{and}
  \bibinfo{author}{\bibfnamefont{B.}~\bibnamefont{{\v{Z}}ek{\v{s}}}},
  \bibinfo{journal}{J. Chem. Phys.} \textbf{\bibinfo{volume}{87}},
  \bibinfo{pages}{5957} (\bibinfo{year}{1987}).

\bibitem[{\citenamefont{Fahrenberger et~al.}(2014)\citenamefont{Fahrenberger,
  Xu, and Holm}}]{FXH:JCP:14}
\bibinfo{author}{\bibfnamefont{F.}~\bibnamefont{Fahrenberger}},
  \bibinfo{author}{\bibfnamefont{Z.}~\bibnamefont{Xu}}, \bibnamefont{and}
  \bibinfo{author}{\bibfnamefont{C.}~\bibnamefont{Holm}}, \bibinfo{journal}{J.
  Chem. Phys.} \textbf{\bibinfo{volume}{141}}, \bibinfo{pages}{064902}
  (\bibinfo{year}{2014}).

\bibitem[{\citenamefont{Hatlo and Lue}(2008)}]{HL:SM:2008}
\bibinfo{author}{\bibfnamefont{M.~M.} \bibnamefont{Hatlo}} \bibnamefont{and}
  \bibinfo{author}{\bibfnamefont{L.}~\bibnamefont{Lue}}, \bibinfo{journal}{Soft
  Matter} \textbf{\bibinfo{volume}{4}}, \bibinfo{pages}{1582}
  (\bibinfo{year}{2008}).

\bibitem[{\citenamefont{Wang and Ma}(2010)}]{WM:JPCB:2010}
\bibinfo{author}{\bibfnamefont{Z.~Y.} \bibnamefont{Wang}} \bibnamefont{and}
  \bibinfo{author}{\bibfnamefont{Y.~Q.} \bibnamefont{Ma}}, \bibinfo{journal}{J.
  Phys. Chem. B} \textbf{\bibinfo{volume}{114}}, \bibinfo{pages}{13386}
  (\bibinfo{year}{2010}).

\bibitem[{\citenamefont{Gan et~al.}(2012)\citenamefont{Gan, Xing, and
  Xu}}]{GXX:JCP:2012}
\bibinfo{author}{\bibfnamefont{Z.}~\bibnamefont{Gan}},
  \bibinfo{author}{\bibfnamefont{X.}~\bibnamefont{Xing}}, \bibnamefont{and}
  \bibinfo{author}{\bibfnamefont{Z.}~\bibnamefont{Xu}}, \bibinfo{journal}{J.
  Chem. Phys.} \textbf{\bibinfo{volume}{137}}, \bibinfo{pages}{034708}
  (\bibinfo{year}{2012}).

\bibitem[{\citenamefont{Jadhao et~al.}(2013)\citenamefont{Jadhao, Solis, and
  de~la Cruz}}]{JSC:JCP:13}
\bibinfo{author}{\bibfnamefont{V.}~\bibnamefont{Jadhao}},
  \bibinfo{author}{\bibfnamefont{F.~J.} \bibnamefont{Solis}}, \bibnamefont{and}
  \bibinfo{author}{\bibfnamefont{M.~O.} \bibnamefont{de~la Cruz}},
  \bibinfo{journal}{J. Chem. Phys.} \textbf{\bibinfo{volume}{138}},
  \bibinfo{pages}{054119} (\bibinfo{year}{2013}).

\bibitem[{\citenamefont{Zwanikken and {de la Cruz}}(2013)}]{ZD:PNAS:13}
\bibinfo{author}{\bibfnamefont{J.~W.} \bibnamefont{Zwanikken}}
  \bibnamefont{and} \bibinfo{author}{\bibfnamefont{M.~O.} \bibnamefont{{de la
  Cruz}}}, \bibinfo{journal}{Proc. Nat. Acad. Sci. USA}
  \textbf{\bibinfo{volume}{110}}, \bibinfo{pages}{5301} (\bibinfo{year}{2013}).

\bibitem[{\citenamefont{Bakhshandeh et~al.}(2011)\citenamefont{Bakhshandeh, dos
  Santos, and Levin}}]{Bakhshandeh:PRL:2011}
\bibinfo{author}{\bibfnamefont{A.}~\bibnamefont{Bakhshandeh}},
  \bibinfo{author}{\bibfnamefont{A.~P.} \bibnamefont{dos Santos}},
  \bibnamefont{and} \bibinfo{author}{\bibfnamefont{Y.}~\bibnamefont{Levin}},
  \bibinfo{journal}{Phys. Rev. Lett.} \textbf{\bibinfo{volume}{107}},
  \bibinfo{pages}{107801} (\bibinfo{year}{2011}).

\bibitem[{\citenamefont{Cai et~al.}(2007)\citenamefont{Cai, Deng, and
  Jacobs}}]{CDJ:JCP:2007}
\bibinfo{author}{\bibfnamefont{W.}~\bibnamefont{Cai}},
  \bibinfo{author}{\bibfnamefont{S.}~\bibnamefont{Deng}}, \bibnamefont{and}
  \bibinfo{author}{\bibfnamefont{D.}~\bibnamefont{Jacobs}},
  \bibinfo{journal}{J. Comput. Phys.} \textbf{\bibinfo{volume}{223}},
  \bibinfo{pages}{846} (\bibinfo{year}{2007}).

\bibitem[{\citenamefont{Gan and Xu}(2011)}]{GX:PRE:2011}
\bibinfo{author}{\bibfnamefont{Z.}~\bibnamefont{Gan}} \bibnamefont{and}
  \bibinfo{author}{\bibfnamefont{Z.}~\bibnamefont{Xu}}, \bibinfo{journal}{Phys.
  Rev. E} \textbf{\bibinfo{volume}{84}}, \bibinfo{pages}{016705}
  (\bibinfo{year}{2011}).

\bibitem[{\citenamefont{Messina}(2002)}]{Messina:JCP:2002}
\bibinfo{author}{\bibfnamefont{R.}~\bibnamefont{Messina}}, \bibinfo{journal}{J.
  Chem. Phys.} \textbf{\bibinfo{volume}{117}}, \bibinfo{pages}{11062}
  (\bibinfo{year}{2002}).

\bibitem[{\citenamefont{Tanaka}(2003)}]{Tanaka:PRE:2003}
\bibinfo{author}{\bibfnamefont{M.}~\bibnamefont{Tanaka}},
  \bibinfo{journal}{Phys. Rev. E} \textbf{\bibinfo{volume}{68}},
  \bibinfo{pages}{061501} (\bibinfo{year}{2003}).

\bibitem[{\citenamefont{Diehl and Levin}(2006)}]{DL:JCP:2006}
\bibinfo{author}{\bibfnamefont{A.}~\bibnamefont{Diehl}} \bibnamefont{and}
  \bibinfo{author}{\bibfnamefont{Y.}~\bibnamefont{Levin}}, \bibinfo{journal}{J.
  Chem. Phys.} \textbf{\bibinfo{volume}{125}}, \bibinfo{pages}{054902}
  (\bibinfo{year}{2006}).

\bibitem[{\citenamefont{Lenz and Holm}(2008)}]{LH:EPJE:2008}
\bibinfo{author}{\bibfnamefont{O.}~\bibnamefont{Lenz}} \bibnamefont{and}
  \bibinfo{author}{\bibfnamefont{C.}~\bibnamefont{Holm}},
  \bibinfo{journal}{Eur. Phys. J. E.} \textbf{\bibinfo{volume}{26}},
  \bibinfo{pages}{191} (\bibinfo{year}{2008}).

\bibitem[{\citenamefont{Shklovskii}(1999)}]{Shklovskii:PRE:1999}
\bibinfo{author}{\bibfnamefont{B.~I.} \bibnamefont{Shklovskii}},
  \bibinfo{journal}{Phys. Rev. E} \textbf{\bibinfo{volume}{60}},
  \bibinfo{pages}{5802} (\bibinfo{year}{1999}).

\bibitem[{\citenamefont{Quesada-P\'erez
  et~al.}(2003)\citenamefont{Quesada-P\'erez, Gonz\'alez-Tovar,
  Mart\'in-Molina, Lozada-Cassou, and Hidalgo-\'Alvarez}}]{QGM+:CPC:2003}
\bibinfo{author}{\bibfnamefont{M.}~\bibnamefont{Quesada-P\'erez}},
  \bibinfo{author}{\bibfnamefont{E.}~\bibnamefont{Gonz\'alez-Tovar}},
  \bibinfo{author}{\bibfnamefont{A.}~\bibnamefont{Mart\'in-Molina}},
  \bibinfo{author}{\bibfnamefont{M.}~\bibnamefont{Lozada-Cassou}},
  \bibnamefont{and}
  \bibinfo{author}{\bibfnamefont{R.}~\bibnamefont{Hidalgo-\'Alvarez}},
  \bibinfo{journal}{Chem. Phys. Chem.} \textbf{\bibinfo{volume}{4}},
  \bibinfo{pages}{234} (\bibinfo{year}{2003}).

\bibitem[{\citenamefont{Martin-Molina et~al.}(2003)\citenamefont{Martin-Molina,
  Quesada-P{\'e}rez, Galisteo-Gonz{\'a}lez, and Hidalgo-Alvarez}}]{MQG:JPCM:03}
\bibinfo{author}{\bibfnamefont{A.}~\bibnamefont{Martin-Molina}},
  \bibinfo{author}{\bibfnamefont{M.}~\bibnamefont{Quesada-P{\'e}rez}},
  \bibinfo{author}{\bibfnamefont{F.}~\bibnamefont{Galisteo-Gonz{\'a}lez}},
  \bibnamefont{and}
  \bibinfo{author}{\bibfnamefont{R.}~\bibnamefont{Hidalgo-Alvarez}},
  \bibinfo{journal}{J. Phys. Condens. Matter} \textbf{\bibinfo{volume}{15}},
  \bibinfo{pages}{S3475} (\bibinfo{year}{2003}).

\bibitem[{\citenamefont{Quesada-P{\'e}rez
  et~al.}(2005)\citenamefont{Quesada-P{\'e}rez, Gonz{\'a}lez-Tovar,
  Mart{\'\i}n-Molina, Lozada-Cassou, and Hidalgo-{\'A}lvarez}}]{QGM:CSA:05}
\bibinfo{author}{\bibfnamefont{M.}~\bibnamefont{Quesada-P{\'e}rez}},
  \bibinfo{author}{\bibfnamefont{E.}~\bibnamefont{Gonz{\'a}lez-Tovar}},
  \bibinfo{author}{\bibfnamefont{A.}~\bibnamefont{Mart{\'\i}n-Molina}},
  \bibinfo{author}{\bibfnamefont{M.}~\bibnamefont{Lozada-Cassou}},
  \bibnamefont{and}
  \bibinfo{author}{\bibfnamefont{R.}~\bibnamefont{Hidalgo-{\'A}lvarez}},
  \bibinfo{journal}{Colloids and Surfaces A: Physicochemical and Engineering
  Aspects} \textbf{\bibinfo{volume}{267}}, \bibinfo{pages}{24}
  (\bibinfo{year}{2005}).

\bibitem[{\citenamefont{Schneider et~al.}(2011)\citenamefont{Schneider,
  Hanisch, Wedel, Jusufi, and Ballauff}}]{SHW:JCIS:11}
\bibinfo{author}{\bibfnamefont{C.}~\bibnamefont{Schneider}},
  \bibinfo{author}{\bibfnamefont{M.}~\bibnamefont{Hanisch}},
  \bibinfo{author}{\bibfnamefont{B.}~\bibnamefont{Wedel}},
  \bibinfo{author}{\bibfnamefont{A.}~\bibnamefont{Jusufi}}, \bibnamefont{and}
  \bibinfo{author}{\bibfnamefont{M.}~\bibnamefont{Ballauff}},
  \bibinfo{journal}{J. Colloid Interf. Sci.} \textbf{\bibinfo{volume}{358}},
  \bibinfo{pages}{62} (\bibinfo{year}{2011}).

\end{thebibliography}

\end{document}